\documentclass[10pt,a4paper,english]{article}\usepackage[]{graphicx}\usepackage[]{xcolor}
\makeatletter
\def\maxwidth{ %
  \ifdim\Gin@nat@width>\linewidth
    \linewidth
  \else
    \Gin@nat@width
  \fi
}
\makeatother

\definecolor{fgcolor}{rgb}{0.345, 0.345, 0.345}

\usepackage{framed}
\makeatletter
 {\par\unskip\endMakeFramed%
 \at@end@of@kframe}
\makeatother

\definecolor{shadecolor}{rgb}{.97, .97, .97}
\definecolor{messagecolor}{rgb}{0, 0, 0}
\definecolor{warningcolor}{rgb}{1, 0, 1}
\definecolor{errorcolor}{rgb}{1, 0, 0}
\newenvironment{knitrout}{}{} 

\usepackage{alltt}

\usepackage[hyphens]{url}
\usepackage{amsmath}
\usepackage{amsfonts}
\RequirePackage{hyperref}
\RequirePackage{ifthen}
\providecommand{\email}[1]{\href{mailto:#1}{\normalfont\texttt{#1}}}
\usepackage{amssymb}
\usepackage{amsthm}
\usepackage{doi}

\theoremstyle{plain}

\theoremstyle{definition}

\theoremstyle{remark}

\PassOptionsToPackage{hyphens}{url}\usepackage{hyperref}
\usepackage{hyperref}
\usepackage[square,numbers]{natbib}
\usepackage{orcidlink}

\usepackage[newitem,newenum,increaseonly]{paralist}

\usepackage[colorinlistoftodos]{todonotes}
\usepackage{skak}
\usepackage{xskak}
\usepackage{chessboard}
\xskakset{style=UF}
\skaklanguage[english]
\storechessboardstyle{shabby}{%
boardfontseries=b}
\usepackage[notheorems]{common}
\usepackage{graphicx} 
\usepackage[figuresleft]{rotating} 

\providecommand{\cfacUL}[2]{\mathUL{\gamma}{#1}{#2}}
\providecommand{\cfac}{\cfacUL{}{}}

\providecommand{\hatvect}[1]{\MATHIT{\hat{\vect{#1}}}}
\providecommand{\hatMtx}[1]{\MATHIT{\hat{\Mtx{#1}}}}
\providecommand{\trbasev}[1][{}]{\vectUL{e}{\trsym}{#1}}

\usepackage{tikz}
\usetikzlibrary{shapes,decorations,arrows,calc,arrows.meta,fit,positioning}
\tikzset{
    -Latex,auto,node distance =1 cm and 1 cm,semithick,
    state/.style ={ellipse, draw, minimum width = 0.7 cm},
    point/.style = {circle, draw, inner sep=0.04cm,fill,node contents={}},
    bidirected/.style={Latex-Latex,dashed},
    el/.style = {inner sep=2pt, align=left, sloped}
}


    \usepackage{enumerate} 
		\usepackage{fancyvrb} 
    \definecolor{orange}{cmyk}{0,0.4,0.8,0.2}
    \definecolor{darkorange}{rgb}{.71,0.21,0.01}
    \definecolor{darkgreen}{rgb}{.12,.54,.11}
    \hypersetup{
      breaklinks=true,  
      colorlinks=true,
      urlcolor=blue,
      linkcolor=darkorange,
      citecolor=darkgreen,
      }

		\let\oldciteauthor=\citeauthor
		\def\citeauthor#1{\hypersetup{citecolor=black}\oldciteauthor{#1}}
    
\IfFileExists{upquote.sty}{\usepackage{upquote}}{}
\begin{document}

\title{Inferring Piece Value in Chess and Chess Variants}
\author{\orcidlink{0000-0002-4197-6195} Steven E. Pav \thanks{\email{steven@gilgamath.com}}}

\maketitle

\begin{abstract}
We use logistic regression to estimate the value of the pieces in standard chess and
several chess variants, namely Chess 960, Atomic chess, Antichess, and Horde chess.
We perform our regressions on several years of data from Lichess, the free and open-source
internet chess server.  \cite{lichess}
We use the published player ratings to control for the confounding effect of differential player skill.
We adjust for the attenuation bias in regressions due to the noise in observed ratings.
We find that major piece values, relative to the value of a pawn, are fairly consistent with historical valuation systems.
However we find slightly higher value to bishops than knights.
We find that piece values are smaller, in absolute value, in Atomic and Antichess than standard chess.
We also present approximate values of the pieces to equalize odds when players of varying skill face off.
We briefly consider self-play experiments using the Stockfish engine, which give a contrasting view of piece value.
\end{abstract}

\section{Introduction}
\label{sec:introduction}

Chess has long fascinated mathematicians and natural scientists. 
The idea that one could construct a machine that played chess
dates back to the ``Mechanical Turk'' of the 18th century. \cite{enwiki:1270044935,standage2002turk}
In the twentieth century, creating a computer program that could play chess
as well as a human was one of the early goals of computer scientists, with
Turing, Shannon, Simon and many other early pioneers taking up the task.
\cite{Turing1988,Shannon1988,Simon1988,Levy1988}
The eventual supremacy of machines over humans was apparent with the defeat of Kasparov by Deep Blue in 1997.
\cite{pandolfini1997kasparov,enwiki:1261068408}

The development of grandmaster-level chess computers did not rely on newly discovered deep mathematical truths about chess,
rather they followed from the increasing availability of computational power.
Deep Blue and its modern kin relied heavily on lookahead, considering the full tree of forward moves and 
picking the move that maximized your worst-case outcome.  \cite{birmingham1988tree}
The introduction of neural nets to chess programs has improved their play, 
but their internal workings are even more opaque than the Mechanical Turk.
\cite{klein2022neuralnetworkschess,chitale2024implementing}

Minimax search can only be performed to some finite number of moves ahead,
eventually relying on static evaluation of the terminal positions if checkmate has not been reached.
Static evaluation often relies on some heuristics, including the balance of pieces,
their relative strengths, the locations of the pieces, whether any pieces are under threat,
whether there are ``passed pawns'' which threaten to promote and so on.
\cite{Levy1988}
This sketch also roughly describes how humans classically were taught to play chess:
one attempts to visualize each possible move and one's opponent's response, out to some
number of moves, with some way to rank the different positions.
The static evaluation that humans use is less precise than that implemented by computers,
but usually includes some accounting of the relative ``value'' of pieces.

In this paper we consider the relative values of the pieces in chess and several chess variants,
for use by both humans and computers.
When first learning chess, the author was taught that a knight had the same value as three pawns,
as did a bishop, that a rook was worth five pawns, and a queen was worth nine.  \cite{winter2012,fine1942chess}

This valuation system is only one of many that have been devised, however, some dating back
to the nineteenth century.  \cite{enwiki:1265673175,kaufman1999evaluation,betza1996}
\citeauthor{maizelis2014soviet}, for example, lists the ``relative strengths of the pieces'' as 3.5, 3.5, 5, and 10 (or 9.5),
values which are determined by their ``mobility in open space''. \cite{maizelis2014soviet}
It is noted, however that piece values are relative to their position on the board, and
piece values may change in the endgame.
\citeauthor{lasker1926} is similarly elusive, giving more than one valuation system in his
book.  \cite[pages 35, 107]{lasker1926}
It is a simple exercise to compute the average number of squares each piece could reach on an empty
board, namely 
knight: 5.25,
bishop: 8.75,
rook: 14,
queen: 22.75, and 
king: 6.5625.
Under this calculation, which was known in the nineteenth century, the pawn is assigned value 1.75. 
\cite{winter2012}

In this paper we estimate piece values by observing the imbalance of pieces in games played between human
players on Lichess, an open source chess server with thousands of users. \cite{lichess}
Lichess publishes games played on the platform, served in monthly bundles.
The most recent bundles for standard chess contain over ninety million games played per month.
Players may also play one of several chess variants on Lichess, including Atomic chess, Antichess, 
Chess 960 (also known as Fisher Random Chess), and several others.  \cite{pritchard2000popular,pritchard2007encyclopedia}
The Lichess database contains over 20 million games played for many of these variants.
We observe outcomes, then use logistic regression to compute ``values'' of the 
pieces.  \cite{gupta2023,gilgamath-atomic-two,gilgamath-atomic-three,mccullagh2019generalized}

There are several ways one could interpret the idea of a ``piece value'' which could be computed from this data:

\begin{compactenum}
\item For a randomly selected game played on Lichess, at some randomly selected moment during the game,
attempt to predict which player will win based purely on the balance of pieces.
While this interpretation of value would be straightforward to estimate, it has some deficiencies: 
some games are played by very unskilled players; 
or at very short time controls;
some games end due to technical issues (\emph{e.g.} a dropped internet connection);
some players cheat;
some games end quickly due to traps, such as Scholar's mate.

\item With some data cleaning we can perhaps estimate a piece value which would allow us to,
for randomly selected skilled and motivated players on Lichess, at some randomly selected moment during a leisurely game,
played to a checkmate or draw, and which doesn't end early via a trap, 
attempt to predict which player will win based purely on the balance of pieces.

While this might be most useful if we were, say, wagering on Lichess games\footnote{We are not.},
such a scoring system might not improve a computer's or a human's static evaluation of positions.
This is because this analysis is purely \emph{observational}, while we would prefer some kind of
\emph{causal} interpretation.  \cite{pearl2018book,angrist2009mostly,hernan2020causal}
There are a number of confounding effects which could degrade the quality of such a valuation when
used for strategic purposes.  

One obvious confounder is player skill. 
For example, the author could not beat Magnus Carlsen in a game even if given a queen advantage in the opening.
What makes skill a potential confounder is that a skilled player can make better use of a piece advantage, 
but also a skilled player is less likely to get themselves into a disadvantageous position in the first place.
If we randomly sample games on Lichess we are likely to select imbalanced positions.

Another problem with our observational setup is that selected positions may not be \emph{quiescent} ones.
Consider the following position from the King's Knight Opening. White holds a material imbalance after
taking black's knight, but the white knight will immediately be recaptured on the next move.
Ignoring the positional differences, the material imbalance is illusory. This position is not quiescent.
Randomly sampling positions in games will catch some positions like this in our sample.
The net effect should be to \emph{understate} the value of a material imbalance.

\newgame
\setchessboard{style=shabby}
\newchessgame[style=shabby,setfen=r1bqkbnr/pppp1ppp/8/4p3/3NP3/3P4/PPP2PPP/RNBQKB1R b KQkq - 0 4,mover=b,moveid=4b]
\mainline{4... pxd4} to recapture.
\setchessboard{style=shabby}
\chessboard[style=shabby,showmover=false]

\item Supposing that we perform additional data cleaning to also remove non-quiescent positions,
and we somehow adjust for confounders, we can perhaps create a score which would be useful for 
\emph{humans} to use in evaluating material imbalance in static evaluation.
Note that this version of piece value value would be based on observing the outcomes from games
played by (imperfect) humans. 
Presumably a computer, or a very skilled human, could better capitalize on a difference in material.
This is perhaps a restatement of the confounding effects of skill.

\item Perhaps if we adjust for imperfect play we can then find a valuation useful for computers.
It seems unlikely that our approach, using Lichess data, can overcome this final hurdle.
Instead we suggest that some kind of ``distillation'' could be performed by running the same
analysis on the outcomes of games played between different (or different instances of the same) computer programs. 
Unfortunately computer programs were historically based on lookahead and static evaluation,
and so the sample might be biased by those valuations. 
To control for this, we should use computer programs built without internal piece values, perhaps
using some variant of AlphaZero, which is built with reinforcement learning.  \cite{silver2017masteringchessshogiselfplay,DBLP:journals/corr/abs-2009-04374}

Note that in such a study, we could also control for the confounding of skill and selected position by
randomly generation positions then observing outcomes. We could also perform ``what if'' experiments
by adding or removing one piece from randomly selected positions, observing outcomes with and without the piece,
to balance the dataset. We leave such a study for further research.
\end{compactenum}

\section{Variants}

In this paper we consider relative piece value in chess and a number of chess variants. 
To distinguish these, we will sometimes refer to chess as ``standard chess.''
For a number of reasons we will focus on variants instead of standard chess, among them:

\begin{compactenum}
\item The monthly game files standard chess in the Lichess database can be quite large.
	Most are greater than 10Gb in size, and in recent years closer to 30Gb in size.
	This is very unwieldy for our data pipeline.
\item As noted elsewhere, many standard chess games played on Lichess end quickly with
	one player falling prey to a trap. This dilutes the value of the data, since trap endings
	have to be identified and ignored.
\item The author is simply more interested in variants.
\end{compactenum}

For these reasons our analysis will focus only on a limited run of standard chess games.
In place, we will consider instead the full database of Chess 960 games.
We believe the valuations in Chess 960 will be applicable to play in standard chess,
except possibly during openings.

\subsection{Chess 960}

Chess 960 is a chess variant that uses the same board, pieces, and rules of standard chess,
but the pieces start from a different (randomly selected) position.  \cite{pritchard2007encyclopedia}
The opening positions must follow some rules (bishops are on opposite colored squares,
the king is between the rooks), which result in 960 different possible opening positions.
We illustrate one below, which is number 512 in the \citeauthor{scharnagl2004fischer} Scheme.
\cite{scharnagl2004fischer} 

\newgame
\newchessgame[setfen=bbrnqknr/pppppppp/8/8/8/8/PPPPPPPP/BBRNQKNR w KQkq - 0 1,moveid=1w]
Chess 960 Opening Position 512.
\chessboard[style=shabby,showmover=false]

In addition to the differences in starting position, the castling rules are slightly more complicated.
We believe the differences to standard chess are small enough that valuations in Chess 960 will be applicable
to standard chess, and that the database will lack traps.

\subsection{Atomic chess}



Atomic chess is another popular variant on Lichess.  \cite[page 42]{pritchard2007encyclopedia}
The game is played with the same board, pieces, piece movements, and opening position as standard chess.
As in standard chess the goal is to checkmate your opponent, or otherwise remove your opponent's king
from the board via an ``explosion''.
These are due to the major change in rules from standard chess:

\begin{compactenum}
\item When you make a capture, the captured piece is removed from the board as in standard chess.
	However, the \emph{capturing piece} is also removed from the board.
	In addition, any non-pawn piece on one of the up to eight adjacent squares is also removed from the board\footnote{On Lichess, the sound of an explosion is issued.}.
\end{compactenum}

We illustrate the effect of an explosive capture below. 
White's capture on \textbf{b7} removes six pieces from the board.

\newgame
\newchessgame[setfen=rnb1kb1r/pp2pppp/Bqp2n2/3p4/8/2N1PQ2/PPPP1PPP/R1B1K1NR w KQkq - 0 5,moveid=5w]
\chessboard[tinyboard,showmover=false]
\mainline{5. Bxb7}
\newchessgame[setfen=4kb1r/p3pppp/2p2n2/3p4/8/2N1PQ2/PPPP1PPP/R1B1K1NR b KQkq - 0 5,moveid=5b]
\chessboard[tinyboard,showmover=false]

The explosion rule has a number of interesting consequences. Among them: 
the king cannot capture, as to do so would be suicidal;
kings may occupy adjacent squares (in which case neither can be put in check);
a lone queen, for example, can cause checkmate on the edge of the board;
one can ignore check or put oneself in check by causing a regicidal explosion.
We illustrate this possibility in the following artificial position from 
which black can prevail by taking the pawn on \textbf{d4}.

\newgame
\newchessgame[setfen=3q3k/1pp1ppb1/r1n2n2/1B1p1Kp1/p2P1P2/N3P1P1/PPP4P/R1B4R b - - 0 12,moveid=12b,result=0--1]
\mainline{12... Nxd4!!#}
\newgame
\newchessgame[setfen=3qk3/1pp1ppb1/r1n2n2/1B1p2p1/p2P1P2/N1K1P1P1/PPP4P/R1B4R b - - 0 12,moveid=12b,result=0--1]
\chessboard[showmover=true]

Because it is relatively harder to remove pawns from the board, one suspects they hold a greater relative value
in this variant, although they can more easily get blocked by each other.
The following position is a win in 10 for black. But remove the pawn at \textbf{c5} and the position is a dead draw,
as none of the pawns can possibly be removed from the board.
Passed pawns can be particularly valuable, as they can promote to a queen, which can give mate unassisted.

\newgame
\newchessgame[setfen=5k2/8/6p1/1pp3P1/7p/pP5P/P7/6K1 b - - 0 21,moveid=21b]
\chessboard[showmover=true]

\subsection{Antichess}

Antichess, sometimes referred to as ``Losing Chess,''
is a fairly old variant, with a number of sub-variants.  \cite{pritchard2000popular,andrejic2018ultimate}
Antichess is played with the same board, pieces, piece movements, and opening position as standard chess.
The goal of Antichess, however, is to arrive at a position where it is your turn and
you have no valid moves, either because you have lost all your pieces, or you have stuck pawns.
To make such an ending more likely, the rules of standard chess are changed in the following ways:

\begin{compactenum}
\item If you can make a capture on your turn, you must make a capture. If there is more than one, you can
	choose among them. If you can make only one capture, you must make that capture. 
	Note that this has the effect of greatly reducing the branching degree in many positions, and makes far
	lookahead possible for both computers and humans.
\item The king loses all royal powers. As such, check is not an active concern, the king may be captured like any other piece,
	castling is not allowed, one can promote to a king and have multiple kings on the board\footnote{Another variant called, ``Loser's Chess'' does not make this change, and keeps royal powers.}.
\end{compactenum}

Antichess games can end in a draw, either by mutual agreement (often when each player has exactly one king on the board),
or when neither player can possibly prevail (most often when each player holds exactly one bishop on opposite colored squares).

One interesting fact about Antichess is that it is a solved game in theory!  
Because of the reduced search space caused by obligatory captures, 
\citeauthor{watkins2017losing} performed a massive computer search to establish that 
\newgame
\mainline{1. e3} is a winning move for white.
\cite{watkins2017losing}
The lookahead required is beyond the scope of human players, however, 
and the game is interesting enough that one sees over 
300 thousand games played a month on Lichess.
Moreover, it is not the case that players uniformly choose this winning opening move.

In Antichess, pieces will be shown to have \emph{negative} value, although 
we believe that position has a much greater impact on static valuation than material.
Indeed, one can easily rearrange positions to completely change the balance of the game.
For example, while \newgame \mainline{1. e3} is theoretically a win for white, 
\newgame
\mainline{1. d3} is an easy win for black!


\subsection{Horde}

Horde chess is a variant with an asymmetrical starting position. 
\cite{enigma2021definitive}\cite[page 90]{pritchard2007encyclopedia}
White starts with 36 pawns arranged as in the following board, 
while black uses the full complement of pieces in their usual arrangement. 
White wins by checkmating, black wins by eliminating all of white's pieces.
The white pawns on the first or second rank may move one or two squares
for their first move. Black may castle under the usual rules.

\newchessgame
\def\hordefen{rnbqkbnr/pppppppp/8/1PP2PP1/PPPPPPPP/PPPPPPPP/PPPPPPPP/PPPPPPPP}
The Horde opening position.
\chessboard[setfen=\hordefen]

It is not clear that pawn imbalance is meaningful in Horde, and we will mostly
focus on the value of black pieces.

\section{Player Skill}

As noted above, results based on an observational study may lack a causal interpretation because of the
confounding effects of skill. 
Consider the following causal structure:

\vspace{2mm}
\begin{tikzpicture}
		\node[state,dashed] (skill) at (1.7,1.3) {skill};

    \node[state] (position) at (0,0) {position};
    \node[state] (outcome) at (4,0) {outcome};

    \path (position) edge[bend left=0] (outcome);
    \path (skill) edge[dashed,bend left=0] (position);
    \path (skill) edge[dashed,bend right=0] (outcome);
\end{tikzpicture}
\vspace{2mm}

Here the difference in latent skill of the players leads to the observed position,
where we measure the difference in material. The difference in position and the 
skill differences then lead to the observed outcome.

Unfortunately, as hinted in the graph, we do not observe skill directly.
However, we have observed the outcomes of previous games.
We can aggregate the results of previous games to compute a kind of ``skill score''
for each player, which is then used as a \emph{proxy} for the skill of each player,
and then adjust for the effects of skill.
Thankfully such a skill score is computed by Lichess and published in their data.
Thus the causal picture is as follows:

\vspace{2mm}
\begin{tikzpicture}
		\node[state,dashed] (skill) at (3.7,1.3) {skill};

    \node[state] (position) at (2,0) {position};
    \node[state] (outcome) at (6,0) {outcome};
		\node[state] (previous) at (-1.0,1.3) {previous games};
		\node[state] (elo) at (-1.0,0) {skill score};

    \path (position) edge[bend left=0] (outcome);
    \path (skill) edge[dashed,bend left=0] (position);
    \path (skill) edge[dashed,bend right=0] (outcome);
    \path (skill) edge[dashed,bend right=0] (previous);
    \path (previous) edge (elo);
\end{tikzpicture}
\vspace{2mm}

The \emph{Elo rating}, named for its inventor Arpad Elo, was the dominant such skill score in chess for many years. \cite{elo1967proposal,elo1978rating}
While the International Chess Foundation, FIDE, computes rankings based on Elo, Lichess
computes and publishes a different skill score, called ``Glicko-2''.  \cite{glickman1995glicko}
Both skill scores share the following calibration: if two players, with skill scores $s_1, s_2$ play each other,
conditioning on an outright victory (that is, ignoring draws), the probability that player 1 wins is $p_1$ with
\begin{equation}
\label{eqn:elo_calibration}
\frac{p_1}{1 - p_1} = 10^{\wrapParens{s_1 - s_2} / 400}.
\end{equation}
Put in words, a 400 difference in Elo (or Glicko-2) corresponds to a 10 fold difference in odds of prevailing.
We can also write
\begin{equation}
	\ln{\frac{p_1}{1 - p_1}} = \frac{\ln{10}\wrapParens{s_1 - s_2}}{400} = \cfac\wrapParens{s_1 - s_2},
\end{equation}
where $\cfac = \ln{10}/400 \approx 0.0058$.
Note that in this framework, the absolute skill score has no interpretation.

Both of the skill scores mentioned are computed in roughly the following manner.
\begin{compactenum}
\item When a new player is introduced to the pool, they are assigned some default value, usually 1500.
	They may also be assigned a kind of uncertainty value.
\item When two rated players play a match, the observed outcome is used to update the scores of each player,
	potentially along with their uncertainty values.
\end{compactenum}
The update formula for the Elo rating is simple enough that it can be carried out by hand or with just a calculator.

These ratings systems can be viewed as implementing something like the \emph{Extended Kalman Filter}.  \cite{fujii2013extended,szczecinski2021simplifiedkalmanfilteronline}
Before observing a match outcome, one has a noisy estimate of scores satisfying \eqnref{elo_calibration},
along with some quantification of the measurement noise.
One also has a model of how the true score may have changed since the last estimate, for example through training or
age-related decline in abilities. 
The EKF then provides a way to update the estimated scores and uncertainty.

To check the calibration of the Glicko-2 ratings, we consider here standard games played on Lichess in 2013 and 2014.
We restrict our attention to games where both players have at least 50 total games played in the database in the 
(up to) 6 previous months of data.
We also filter on games which ended by a normal win lose or draw, instead of time-out,
and we filter on games with at least 5 minutes allotted to each player.
We also filter on games where the ratings of both players, prior to the match, are at least 1200.
This results in a data set of 4,106,242 games.

We bin the games by the differences in ratings, as computed prior to the game.
We compute the average empirical probability that the underdog, defined as the player with lower pre-game
rating (or black, if they share the same rating), wins the game. 
We consider a draw half a win for purposes of computing these probabilities.
We then plot these empirical probabilities against those obtained from
\eqnref{elo_calibration} on the bin median difference in ratings,
in \figref{elo_I_plot}.

\begin{knitrout}\small
\definecolor{shadecolor}{rgb}{0.969, 0.969, 0.969}\color{fgcolor}\begin{figure}[h]
\includegraphics[width=0.975\textwidth,height=0.691\textwidth]{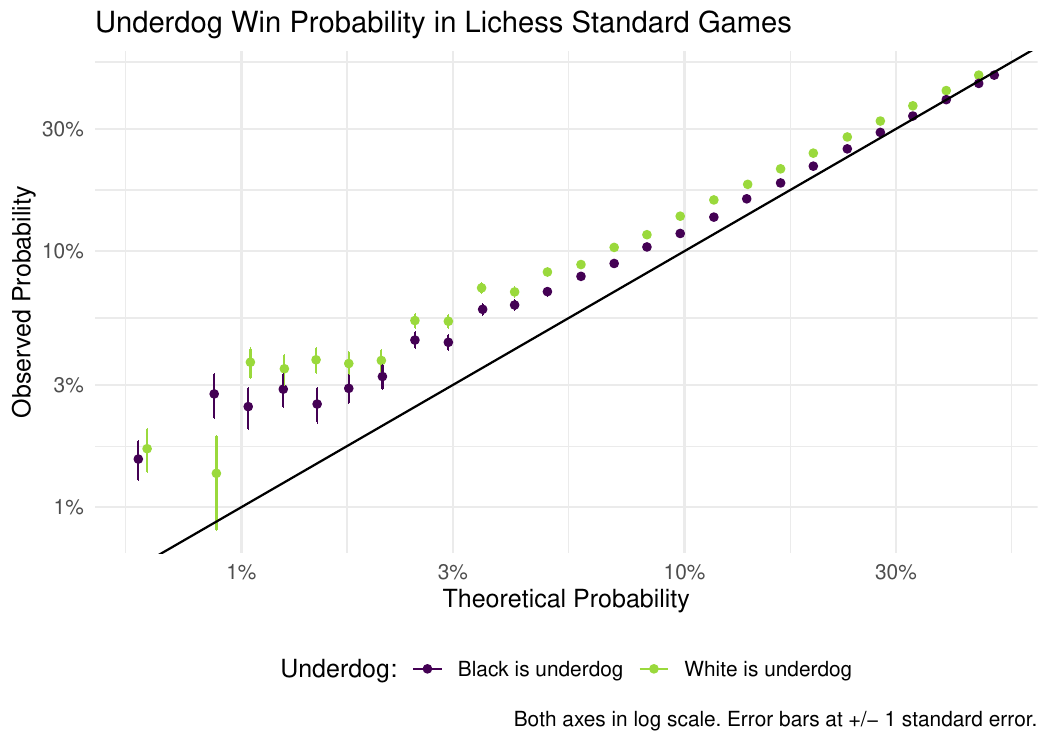} \caption[The empirical probability of a win is plotted against the theoretical probability based on published rating differences for standard Lichess games]{The empirical probability of a win is plotted against the theoretical probability based on published rating differences for standard Lichess games. To show detail, we plot in log scale from the underdog's point of view.}\label{fig:elo_I_plot}
\end{figure}

\end{knitrout}

Two things are apparent from the plot:
\begin{compactenum}
\item White enjoys a small advantage, with slightly higher probability of winning as underdog than black.
	This is the benefit of having the first move and maintaining tempo.
\item The rating differences are not well calibrated: the empirical probability of an underdog victory is higher than
	suggested by the difference in ratings.
\end{compactenum}

Note that there a number of possible explanations of the underdog bias we observe here.
Most of the games in this dataset are played between players with near equal rating;
fully 91.3\% are between players with rating differences
not exceeding 300 points. Perhaps underdogs elect to play against nominally stronger players because they 
know their rating is an underestimate.

A more mathematically tractable explanation is that the ratings themselves are estimates of skill and have measurement error.
Even if this measurement error is unbiased we can expect the empirical win probability to be biased due to 
Jensen's inequality, because the response via a logistic function is nonlinear.
To better quantify this measurement error, and the impact it will have on our analysis, we introduce the regression
formulation we will use in the paper.

\subsection{Regression Analysis}
\label{sec:regression_analysis}

We can view \eqnref{elo_calibration} as one simple version of the kind of equation we will use to model
outcomes of games. A more elaborate one might look as follows:
\begin{equation}
	\label{eqn:outcome_reg_I}
	\log{\frac{p_1}{1 - p_1}} = \beta_0 
	+ \beta_s \wrapParens{s_1 - s_2} 
	+ \beta_R \wrapParens{R_1 - R_2}
	+ \beta_Q \wrapParens{Q_1 - Q_2}
	+ \ldots
\end{equation}
Here, we write $R_i$ as the number of rooks held by player $i$ at the time we snapshot the position,
and $Q_i$ is the number of queens held by player $i$. 
We omit the knights, bishops and pawns here just for expositional clarity, but we will include
them when we perform the actual regressions.
We can estimate the $\beta_j$ coefficients in \eqnref{outcome_reg_I} by 
performing a \emph{logistic regression} on some observed data that includes the measured skill scores,
material differences and outcomes.  \cite{mccullagh2019generalized}
If we encode the data such that player 1 is always white, then $\beta_0$ can be interpreted as the
first player advantage.

If our skill scores are properly calibrated we still expect to find $\beta_s = \cfac$.
However we do not observe latent skill. 
Instead we observe Glicko-2 ratings $\hat{s}_i$ which are a noisy estimate of skill.
If we perform logistic regression on the formulation
\begin{equation*}
\frac{p_1}{1 - p_1} = \beta_0 
	+ \beta_s \wrapParens{\hat{s}_1 - \hat{s}_2},
\end{equation*}
we expect to find $\abs{\beta_s} < \abs{\cfac}$ due to the \emph{attenuation bias}.
\cite{kennedy2003guide}
To see why, think of the limiting situation where the noise in $\hat{s}_i$ is very large;
in this case one expects the estimated $\beta_s$ to be very close to zero, since
the noise is not predictive of outcomes.

\begin{table}[!htbp] \centering 
  \caption{Logistic regression on the standard chess data sample. Standard errors are omitted.} 
  \label{tab:elo_I_regs} 
\small 
\begin{tabular}{@{\extracolsep{5pt}}lc} 
\\[-1.8ex]\hline 
\hline \\[-1.8ex] 
 & \multicolumn{1}{c}{\textit{Dependent variable:}} \\ 
\cline{2-2} 
\\[-1.8ex] & Game Outcome \\ 
\hline \\[-1.8ex] 
 Rescaled Glicko 2 & 0.853 \\ 
  First Player Advantage & 0.071 \\ 
 \hline \\[-1.8ex] 
num observations & 4,106,242 \\ 
\hline 
\hline \\[-1.8ex] 
\end{tabular} 
\end{table}

If we perform such a logistic regression on our small filtered sample of standard chess games,
we get the results shown in \tabref{elo_I_regs}.
We have rescaled the $\hat{s}_i$ by $\cfac$ so the fit coefficient would ideally be 1;
however we observe a fit value of 0.853. 
Our estimate of $\beta_s$ is attenuated by the error in the measurement of the dependent variable.

Note that in this paper, 
when we perform the logistic regression, we will generally omit the standard errors and p-values.
There are a few reasons for this:
\begin{compactenum}
\item 
	When we perform the regression, we encode a win or loss as two identical outcomes of two coin flips,
and encode a draw as two different outcomes in coin flips. 
Thus the regression code is effectively being told there are twice as many independent observations as actually observed matches,
and the reported standard errors would be too small by a factor of $\sqrt{2}$.
\item 
	We will use simulated extrapolation to deal with the noise in our skill proxy. 
It is not clear how to adjust standard errors for this procedure.
\item 
	Given the sample sizes the estimated standard errors will generally be very small, but the regression estimates
may change by many such standard errors if we change the filters we apply on the sample.
For example, including or excluding games played at a one minute time control may change the estimated piece values
by a small amount, but one which is much bigger than the estimated standard errors.
\end{compactenum}

By a somewhat circuitous route we can estimate the noise in the $\hat{s}_i$ measurements.  
We do this via two groups of regressions:

\begin{compactenum}
\item 
	First we construct a large dataset where we know the true deltas $s_1 - s_2$, 
	but add unbiased Gaussian noise with standard deviation $\sigma_s$ to the $s_1 - s_2$ to arrive at the delta $\hat{s}_1 - \hat{s}_2$.
	We construct the true underlying differences $s_1 - s_2$ by sampling from the observed $\hat{s}_1 - \hat{s}_2$, 
	shrunk slightly so that when the noise is added back the sample standard deviations match the observed.
	We sample random outcomes using the properly calibrated law in \eqnref{elo_calibration}, but then perform a regression against the observed $\hat{s}_1 - \hat{s}_2$,
	and compute a logistic regression.
	We then look at the regression coefficient divided by \cfac as a function of $\sigma_s$. 
	The dependent and independent variables are generated by us, so we call these Monte Carlo regressions.
\item 
	Secondly we take our dataset of 4,106,242 games and then add additional unbiased Gaussian noise with standard deviation $\sigma_z$ to the
	$\hat{s}_1 - \hat{s}_2$. 
	We perform the logistic regression against the extra-noisy version of $\hat{s}_1 - \hat{s}_2$,
	again normalizing by \cfac.
	We call these the simulation extrapolation (or SIMEX) regressions.  \cite{carroll1995measurement,doi:10.1177/1536867X0400300407}
	If the original deltas $\hat{s}_1 - \hat{s}_2$ are polluted by noise with standard deviation $\sigma_0$, then the total standard deviation
	after the additional noise should be $\sqrt{\sigma_0^2 + \sigma_z^2}$.
\end{compactenum}

We plot the results of these experiments in \figref{elo_I_weirdness}.
We have used $\sigma_0=58$ to plot the $x$ of the SIMEX experiments,
a value we found via trial and error.
This is all very suggestive that the standard deviation of the noise in differences of standard chess Glicko-2 values is around 58.
Note that this value is about what one would expect by analyzing Elo-like scores from first principles, see
\secref{elo_se} in the Appendix.

\begin{knitrout}\small
\definecolor{shadecolor}{rgb}{0.969, 0.969, 0.969}\color{fgcolor}\begin{figure}[h]
\includegraphics[width=0.975\textwidth,height=0.691\textwidth]{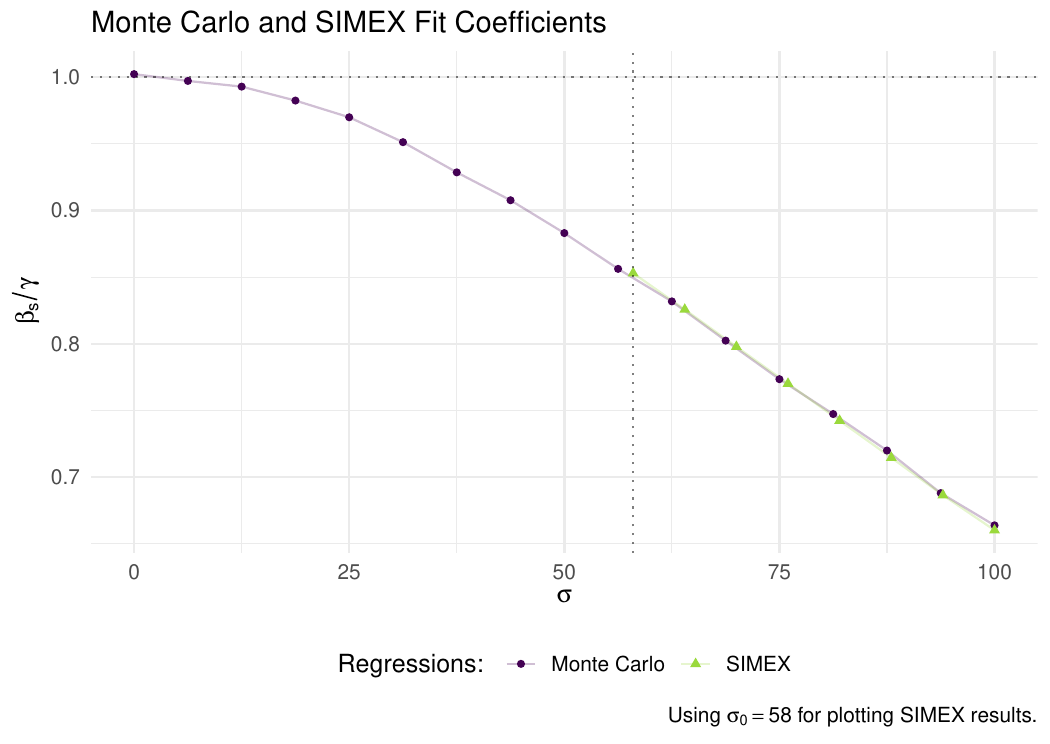} \caption[The ratio of fit coefficient to \cfac is plotted for the Monte Carlo and SIMEX experiments]{The ratio of fit coefficient to \cfac is plotted for the Monte Carlo and SIMEX experiments. The SIMEX data are plotted using $\sigma_0=58$.}\label{fig:elo_I_weirdness}
\end{figure}

\end{knitrout}

\subsection{Extended SIMEX Regressions}

While this calibration exercise is relatively simple, 
we wish to consider more complicated regression formulations,
for instance ones like \eqnref{outcome_reg_I}, but with terms for all the pieces.
It is not immediately clear how the noise in $\hat{s}_1 - \hat{s}_2$ will
affect the estimated piece value coefficients.
That is, above we observed that noise in the ratings causes attenuation in the
coefficient associated with the ratings.
However, the noise in the ratings will likely affect the estimated regression
coefficients for other terms as well.
To examine this effect, we will perform some more Monte Carlo regressions
where we know the exact values of the true regression coefficients.
We will then add varying amounts of noise to the observed rating difference,
performing the regressions, then plotting the fit regression coefficients
as a function of the added noise.

We perform a series of Monte Carlo experiments to examine the effect of attenuation bias
on white's first player advantage and the piece value of a knight.
We take the difference in ratings from the sample of 4,106,242 games considered
in the previous section, but scale them down slightly to match the observed sample standard deviation
if we were to add noise with standard deviation 58.
We then spawn a pseudo position with a differential in the number of knights. 
The difference in knights is randomly sampled as
$$
\Delta N = N_1 - N_2 \sim -2 + \mathcal{B}\wrapParens{4, \frac{\cfac}{2} \Delta s}.
$$
That is, we draw from a binomial distribution on 4 trials with probability equal to one half of $\cfac \Delta s$,
which is the rating effect, then subtract 2 to get the knight imbalance.
Having observed $\Delta N$ we then construct the odds that white wins as:
\begin{equation}
	\label{eqn:outcome_reg_mc_IV}
	\log{\frac{p_1}{1 - p_1}} = 
  \beta_w
	+ \beta_s \Delta s
	+ \beta_N \Delta N.
\end{equation}
We then sample one Bernoulli trial with this probability.
We perform logistic regression against the observed knight differential, 
the noisy rating difference $\Delta \hat{s}$, and an intercept term.
We repeat this several times and compute the median regression coefficients
over the trials for each term.
We repeat this experiment for different values of the noise added to the rating difference, $\sigma$.
In these experiments we use $\beta_w = 15\cfac$, $\beta_N = 45\cfac$ and $\beta_s = \cfac$.
That is, white's first player advantage is equal to $15$ points difference in Glicko-2,
and a single knight is worth $45$.

\begin{knitrout}\small
\definecolor{shadecolor}{rgb}{0.969, 0.969, 0.969}\color{fgcolor}\begin{figure}[h]
\includegraphics[width=0.975\textwidth,height=0.691\textwidth]{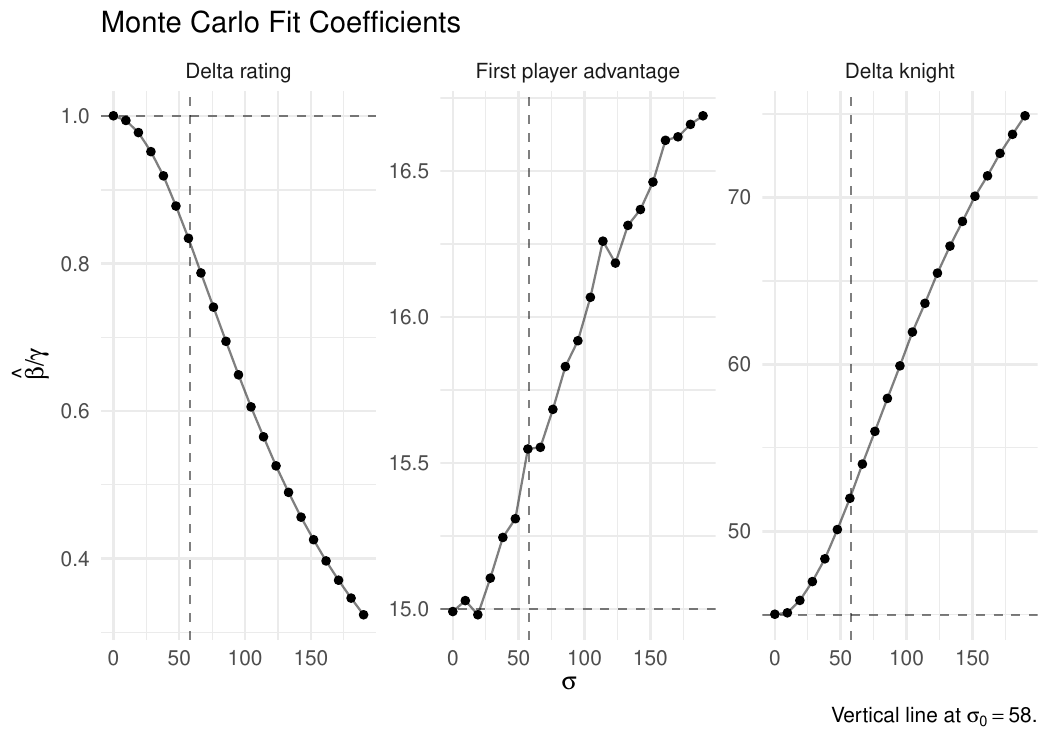} \caption[The logistic regression fit values for the knight Monte Carlo experiments are shown as a function of the added noise]{The logistic regression fit values for the knight Monte Carlo experiments are shown as a function of the added noise. Horizontal lines are plotted at the actual known coefficient values.}\label{fig:elo_IV_mc_plots}
\end{figure}

\end{knitrout}

In \figref{elo_IV_mc_plots} we plot the estimated regression coefficients for the different levels of added noise.
As above, in \figref{elo_I_weirdness} for example,
we see an attenuation bias where the coefficient for the rating difference is diminished by the noisy observation of ratings.
The fit coefficients for first player and difference in knight counts are \emph{also} affected by the noise, even
though their corresponding independent variables are observed without noise.

\subsection{Extrapolation Methodology}

In our actual study of piece values we will observe the regression coefficients for $\sigma = \sigma_0$.
By means of SIMEX, we can add \emph{more} noise and observe regression coefficients for $\sigma > \sigma_0$.
Our job will be to infer the regression coefficients at $\sigma=0$.
That is, we observe the curves to the right of the vertical lines in \figref{elo_IV_mc_plots},
but need to extrapolate back to the zero noise intercept.
We will do this via \emph{regression calibration}. See \secref{reg_cal} in the appendix for more details.

First, for the rating effect, $\beta_s$, we performed a regression of
\begin{align*}
\frac{1}{\hat{\beta}_s} &\sim b_0 + b_2 \sigma^2 + b_4 \sigma^4+ b_6 \sigma^6,
\end{align*}
where we use only the actual and SIMEX regressions with $\sigma \ge \sigma_0$ to get the $\hat{\beta}_s$ and $\sigma$.
Note this deviates from the theoretical exposition, which includes only the intercept and $\sigma^2$ term.
We find here that including the extra terms gives a better extrapolation fit.

For the first player and knight effect coefficients, we performed regressions of
\begin{align*}
	\hat{\beta}_w &\sim c_0 + c_1 \sigma^2\hat{\beta}_s,\\
	\hat{\beta}_N &\sim d_0 + d_1 \sigma^2\hat{\beta}_s,
\end{align*}
again using only data from the SIMEX regressions.
In \figref{elo_IV_mc_extrapolations_plot} we plot the result of these extrapolations,
drawing estimated fit lines for the $0 \le \sigma < \sigma_0$ region,
even though the fit lines were not built using this data. 
We see that the extrapolations do a very good job of fitting the data in the $\sigma < \sigma_0$ range.
The estimated values are given in \tabref{elo_IV_mc_extrapolations_table}.

\begin{table}[ht]
\centering
\begin{tabular}{lrrl}
  \hline
Term & True Value & Estimated Value & Error \\ 
  \hline
Delta rating & 1.000 & 0.994 & 0.55\% \\ 
  First player advantage & 15.000 & 15.118 & 0.79\% \\ 
  Delta knight & 45.000 & 45.092 & 0.20\% \\ 
   \hline
\end{tabular}
\caption{The true and estimated coefficients from the SIMEX and regression calibration experiments are shown.} 
\label{tab:elo_IV_mc_extrapolations_table}
\end{table}

\begin{knitrout}\small
\definecolor{shadecolor}{rgb}{0.969, 0.969, 0.969}\color{fgcolor}\begin{figure}[h]
\includegraphics[width=0.975\textwidth,height=0.691\textwidth]{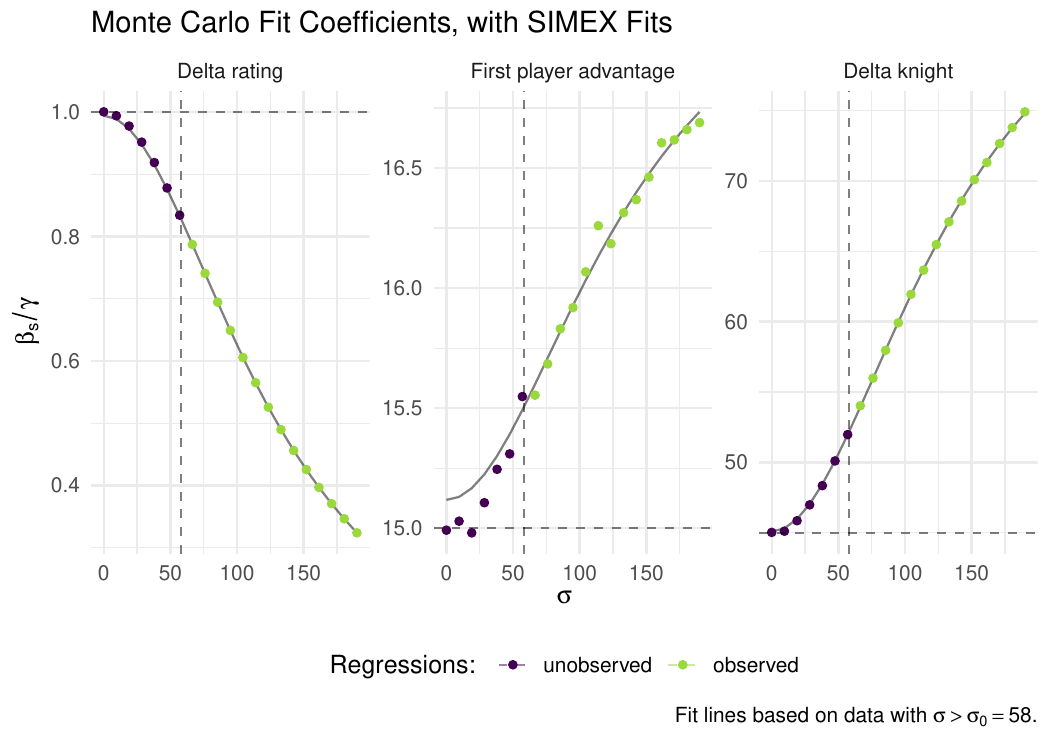} \caption[The logistic regression fit values from \figref{elo_IV_mc_plots} are plotted again, with extrapolation fits using only data with $\sigma \ge \sigma_0=58$]{The logistic regression fit values from \figref{elo_IV_mc_plots} are plotted again, with extrapolation fits using only data with $\sigma \ge \sigma_0=58$. }\label{fig:elo_IV_mc_extrapolations_plot}
\end{figure}

\end{knitrout}

\section{Results}

\subsection{Standard Chess}

As noted above we start from all standard chess games played on Lichess in 2013 and 2014.
We restrict our attention to games where both players have at least 50 total games 
played in the database in the up to 6 previous months of data.
We also filter on games which ended by a normal win lose or draw, instead of time-out,
and we filter on games with at least 5 minutes allotted to each player.
We also filter on games where the ratings of both players, prior to the match, are at least 1200.
This is a total of 
4,106,242 games.
From this set of games we perform the calibration procedure on Glicko-2 scores alone, as described above,
to estimate the $\sigma_0$, the amount of noise in the Glicko-2 ratings.
This we estimate as $\sigma_0=58$.

For the estimation of piece values we perform further filtering on the games.
We filter out games where the total number of plies is less than 10 or greater than 150,
which leaves us with
4,007,111 games.
We then remove games where the recorded piece imbalance seems unusual.
Here this means rejecting cases where the snapshot knight imbalance, bishop imbalance or rook imbalance is greater than 3 in absolute value;
similarly we also reject cases where the absolute queen imbalance is greater than 2 in absolute value.
This results in a set of 
4,006,957 games.
The median game among these lasts to 63 plies.

We perform the SIMEX procedure, then calibration back to the zero noise values as outlined above.
The results are presented in \tabref{standard_simex_I_res_table}.
We repeat this procedure but on filtered subsets of the data.
We first filter out all positions where the next move is a capture.
This results in a set of $3,008,797$ positions.
We then further filter out all positions where the next move after that is a capture,
thus yielding only positions where there is no capture in the next two plies, a set of size
$2,506,200$.

The model we fit includes terms for the rating difference, the first player advantage (``White Advantage''),
a tempo term, and then terms for the differences in number of pawns, knights, bishops, rooks, and queens.
The tempo term is set to one when we snapshot at white's turn, and negative one when we snapshot at black's turn.
Recall that the outcome and material differences are quoted in terms of white's point-of-view, so the
tempo term acts to measure the conditional value of having survived up to snapshot time and having the freedom to make a move.

As seen in \tabref{standard_simex_I_res_table}, the value of all pieces increases upon application of the first filter.
For some pieces the increase is considerable, on the order of 25\% or more.
However, the second filter has very little effect.
Again, our justification for the no-capture filter is that it is a crude way to approximately select only quiescent positions.
However, our goal is to measure whether material imbalance leads to victory; one stepping stone to victory is the capture of 
material, so there is a tension here between capturing quiescent positions and measuring the value of material.
A more sophisticated filter for quiescent positions is warranted, but we leave this for further research.
For the remainder of this paper we will mostly select positions based on the single no-capture filter.

We note that if we consider this regression, and divide the estimated piece values by the estimated pawn values, we
find that 
a knight is worth 2.9 pawns,
a bishop is worth 3.2 pawns,
a rook is worth 4.6 pawns,
and 
a queen is worth 9.6 pawns.

We plot the values, relative to a pawn, in \figref{standard_centipawn_plot_I}, along with the values from
the Chess 960 fits we perform later.
We include dots for the historical valuation systems given in \tabref{value_systems} in the appendix.
This collection of other valuation systems is biased by the easily available scholarship on the matter,
and the density of points does not necessarily indicate consensus view on valuations. 
Our intention in including them is only to illustrate the spread and the approximate values in these systems.
The values we find are roughly in line with these other valuation systems, but for a few notable differences:

\begin{compactenum}
\item 
	We find smaller value for knights than other systems, especially relative to the value of a bishop.
\item 
	We find values on the smaller end of the spectrum for each piece, which perhaps indicates we find a higher valuation for the pawn, which is the numeraire.
\end{compactenum}

We also note that the coefficient associated with rating difference is somewhat smaller than 1, 
even though we performed the calibration procedure, which should correct for the attenuation effect.
The simple interpretation, backed by results we will see later, is that our sample contains some positions
which strongly favor one player or another. For such positions the difference in skills is less relevant.
This effect is more apparent when the snapshot is taken later in the game.

In \tabref{standard_with_confint_table_2} below we repeat the regression results from the procedure applied to the no-capture filtered set
(second column of \tabref{standard_simex_I_res_table})
alongside the na\"{i}ve logistic regression on the sample. 
We see that while the attenuation effect has a relatively large impact on the rating coefficient,
the impact to the other coefficients is actually quite small. 
Moreover, the na\"{i}ve logistic regression
results in \emph{larger} coefficients for the non-rating terms, which compensate for the attenuation on the rating variable.
Ignoring the rating term, the SIMEX and calibration procedure has little impact on the final regression results.

\begin{table}[!htbp] \centering 
  \caption{The estimated coefficients for standard chess are shown for three filtering choices: no filtering based on capture, removing positions with an immediate capture, removing positions with a capture in the next 2 plies.} 
  \label{tab:standard_simex_I_res_table} 
\small 
\begin{tabular}{@{\extracolsep{5pt}}lccc} 
\\[-1.8ex]\hline 
\hline \\[-1.8ex] 
 & \multicolumn{3}{c}{\textit{Dependent variable:}} \\ 
\cline{2-4} 
\\[-1.8ex] & \multicolumn{3}{c}{Game Outcome} \\ 
 & No Filtering & No capture Next 1 & No Capture Next 2 \\ 
\\[-1.8ex] & (1) & (2) & (3)\\ 
\hline \\[-1.8ex] 
 $\Delta$ Glicko 2 & 0.95 & 0.94 & 0.94 \\ 
  White Advantage & 13.60 & 14.40 & 14.00 \\ 
  Tempo & $-$17.60 & 3.20 & $-$5.84 \\ 
  $\Delta$ Pawn & 60.10 & 67.60 & 69.50 \\ 
  $\Delta$ Knight & 147.00 & 195.00 & 199.00 \\ 
  $\Delta$ Bishop & 168.00 & 214.00 & 219.00 \\ 
  $\Delta$ Rook & 248.00 & 311.00 & 316.00 \\ 
  $\Delta$ Queen & 479.00 & 650.00 & 656.00 \\ 
 \hline \\[-1.8ex] 
num observations & 4,006,957 & 3,008,797 & 2,506,200 \\ 
\hline 
\hline \\[-1.8ex] 
\end{tabular} 
\end{table}

\subsubsection{Confidence Intervals}

We briefly consider the question of confidence intervals for these regressions.
That is, how precise are the values in \tabref{standard_simex_I_res_table}?
If we were to examine a different set of games, how different might we expect the results to be?
Can we confidently say that a bishop has higher value than a knight? 

The classical theory of generalized linear models gives confidence intervals for a single logistic regression.
\cite{dobson_introduction_2008}
However, the SIMEX and calibration procedure does not have a standard procedure for confidence intervals,
and instead we fall back on \emph{bootstrapping}.  \cite{efron1994}
The idea is simple: 
we take the sample of 3,008,797 games, and resample from them uniformly \emph{with replacement}.
We then perform the SIMEX and calibration procedure on that sample, record the regression coefficients,
then repeat many times, ideally hundreds of times.
The spread in the recorded regression coefficients across the bootstrap replicates gives some estimate of
the confidence intervals on the regression coefficients.
Unfortunately the SIMEX procedure requires us to recompute a logistic regression many times for each choice of additional noise,
then span a range of additional noises. 
This means each bootstrap replicate requires dozens or even hundreds of logistic regressions on a sample of millions of observations,
and so performing hundreds of bootstrap replications is not practical.

We perform only 100 bootstrap replications. 
From these we compute approximate 95\% confidence intervals on each of the estimated coefficients.
We report these in \tabref{standard_with_confint_table_2}, in the first column.
There we also report the confidence intervals on the na\"{i}ve logistic regression. 
We note that the SIMEX and calibration procedure does not result in appreciably wider confidence intervals than just using logistic regression.
We also note that many of the coefficients we estimated without the no-capture filter (first column of \tabref{standard_simex_I_res_table})
fall far outside the confidence intervals.
Later we will see that when snapshotting games at different points in the game lifetime will also result in coefficients far outside
these intervals.
Thus we conclude that confidence intervals are somewhat misleading in this study, and will not report them.
The sample sizes we use here are sufficiently large that confidence intervals would appear very small, 
but reasonable changes in filters or selection rules will result in changes that break out of those intervals.

Lastly, we note that non-overlapping confidence intervals is not equivalent to a statistically significant difference in coefficients.  \cite{vanBelle2002_STRUTs}
For example, we would like to look at \tabref{standard_with_confint_table_2} and state unequivocally that bishops
have higher value than knights; 
however, our valuation estimates may be negatively correlated, which is to say that we might overestimate bishops when we underestimate knights, on average,
and thus the difference between them is larger than suggested by their marginal confidence intervals.
This is a failing of the method by which we report the confidence intervals, not of the methodology itself;
however, most readers could not easily absorb a table with the full 8 by 8 covariance matrix of the estimates.

\begin{table}[!htbp] \centering 
  \caption{The estimated coefficients for standard chess are shown under filtering for immediate capture. The first column shows coefficients for SIMEX plus calibration, repeating the middle column of \tabref{standard_simex_I_res_table}. The reported numbers in parenthesis are approximate 95\% confidence intervals based on a bootstrap. The second column shows the naive logistic regression with 95\% confidence intervals computed via classical theory.} 
  \label{tab:standard_with_confint_table_2} 
\small 
\begin{tabular}{@{\extracolsep{5pt}}lcc} 
\\[-1.8ex]\hline 
\hline \\[-1.8ex] 
 & \multicolumn{2}{c}{\textit{Dependent variable:}} \\ 
\cline{2-3} 
\\[-1.8ex] & \multicolumn{2}{c}{Game Outcome} \\ 
 & SIMEX + calibration & Raw logistic \\ 
\\[-1.8ex] & (1) & (2)\\ 
\hline \\[-1.8ex] 
 $\Delta$ Glicko 2 & 0.94 & 0.80 \\ 
  & (0.94 - 0.95) & (0.80 - 0.80) \\ 
  & & \\ 
 White Advantage & 14.40 & 14.88 \\ 
  & (13.94 - 14.77) & (14.43 - 15.33) \\ 
  & & \\ 
 Tempo & 3.20 & 3.23 \\ 
  & (2.72 - 3.57) & (2.78 - 3.68) \\ 
  & & \\ 
 $\Delta$ Pawn & 67.60 & 69.27 \\ 
  & (67.05 - 68.00) & (68.75 - 69.79) \\ 
  & & \\ 
 $\Delta$ Knight & 195.00 & 196.80 \\ 
  & (193.10 - 195.70) & (195.60 - 198.00) \\ 
  & & \\ 
 $\Delta$ Bishop & 214.00 & 218.00 \\ 
  & (213.10 - 215.60) & (216.80 - 219.20) \\ 
  & & \\ 
 $\Delta$ Rook & 311.00 & 313.60 \\ 
  & (309.30 - 312.90) & (311.80 - 315.40) \\ 
  & & \\ 
 $\Delta$ Queen & 650.00 & 651.80 \\ 
  & (644.70 - 653.10) & (647.90 - 655.80) \\ 
  & & \\ 
\hline \\[-1.8ex] 
num observations & 3,008,797 & 3,008,797 \\ 
\hline 
\hline \\[-1.8ex] 
\end{tabular} 
\end{table} 

\subsubsection{Expanded terms}

We now expand the formulation and rerun the regression to estimate the value of passed pawns.
A passed pawn is one with no opposing pawn in its file or adjacent files that could prevent it from promoting.
In the following position, white has a passed pawn at \textbf{d5}, while black has one at \textbf{a5}.
Black's pawn on \textbf{b6} is not a passed pawn. 

\newgame
\newchessgame[setfen=5k2/6p1/1p4qp/p1pPp1p1/b1P1Pn2/2P5/2Q3PP/3BB1K1 w - - 0 28,mover=w,moveid=28w]
\chessboard[showmover=false]

We construct dummy variables for the presence of a passed pawn on the 2nd, 3rd or 4th rank (relative to the player's point of view); 
one variable for a passed pawn on the 5th rank; 
one variable for a passed pawn on the 6th rank; 
and one variable for a passed pawn on the 7th rank.
Thus we add 4 independent variables to the regression formulation,
namely the differences in the numbers of such passed pawns, from white's point of view minus those from black's point of view.
We use the same definition of a passed pawn in each of the variants we consider here, 
even when the rules of that particular variant have different implications for how a pawn can block another.
The passed pawn indicator variables do not affect the computation of the pawn indicator variables.
Thus in the delta pawn variable is -1, since white has 6 pawns and black has 7;
the delta in passed pawns at 2,3,4 is -1 since black has one at \textbf{a5} and white has none;
the delta in passed pawns at 5 is 1 since white has one at \textbf{d5} and black has none;
the delta in passed pawns at 6 and at 7 are both zero since there are none on the board for either player.

In \tabref{standard_expand_tab_I}, we list the results of the SIMEX + calibration procedure on the expanded formulation,
in the second column. 
In the first column we repeat the regression results from the smaller regression, 
as given in the second column of \tabref{standard_simex_I_res_table}, for comparison.
The value of a knight and bishop have not changed, while those of the pawn, queen and rook have changed in response
to the extra terms.
As expected, the value of a generic pawn is now depressed, while the value of a ``passer'' is positive and increasing as
they advance to promotion. 
Note that the value of the passed pawns listed here is in addition to the value of an undistinguished pawn,
so a passed pawn on the 7th rank has almost exactly as much value as a knight.

Interpreting piece values relative to that of a pawn is now complicated by the accounting for the value of passed pawns.

\subsubsection{Different snapshots}

We now consider the effect of \emph{when} we snapshot the game.
In chess parlance, \emph{ply} refers to a single piece movement by one player, while a 
\emph{move} (or \emph{full move}) refers to two plies, one by white, one by black.
In the results above we sampled games effectively uniformly at random from the second ply to the last,
but we are also interested in how piece value depends on the phase of the game.
While chess players often divide games into opening, middlegame and endgame,
it is tricky to identify those in an objective, data-driven way, especially at scale.
Instead we use some simple rules to try to capture how value is modulated by game phase.

First we actually snapshotted every eligible game in our dataset at four different places:
uniformly at random, as used above; uniformly among the first one third of the game, defined by ply;
uniformly among the second third; uniformly among the last third.
These do not exactly translate to the phases of a game. 
Moreover, the length of a game is never known in advance to players, so these are not exactly actionable for players.

We tabulate the results of these three regressions in \tabref{standard_expand_tab_I}.
There we see a generally increasing value for pieces during the course of games,
and decreasing value to the difference in rating.
As noted elsewhere, our interpretation is that there are some positions from which the winner is more or less determined,
we are more likely to see these later in a match, and from those positions, rating difference is less predictive of outcome.

We also group games by ranges of ply. For a given range of ply, we look at all four of the snapshots
we took of a game, and select one which is within the ply range, and which satisfies the no-capture condition.
We select at most one snapshot from each game, none if there are none in the ply range which satisfy the filter condition.
We then perform the SIMEX and calibration procedure, and tabulate the results in \tabref{standard_ranged}.
It is important to note that these regressions are conditional on having survived to that given ply.
That is, as similar to the previous time snapshotting pragma, you do not know \emph{a priori} how long a game will be,
so you do not know if piece value at, say, ply 70 will be relevant to you.

The values of the pieces are plotted in \figref{standard_ranged_fit_plot} versus the mean ply of each slice of data.
There the raw values are plotted, as well as the values relative to the estimated value of a pawn.
This is somewhat complicated in this formulation because the pawn value does not include the bonus to passed pawns.
A few notable stories emerge from this plot and the results in \tabref{standard_ranged}:

\begin{compactenum}
\item The value of rating difference is approximately 1.0 for early ply, which we expect due to how calibration works,
	but decreases as the game progresses.
	Again we attribute this to progression to board states which are strongly advantageous to one of the players.
\item White's first player advantage shrinks as the game progresses, as one would expect. It is effectively zero by the 60th ply,
	assuming black can survive until then.
\item The value of Tempo fluctuates through the game. More likely than not this term would have large standard errors,
	and high variance inflation with other terms.
\item The raw value of the pieces is increasing through the game. As a game progresses and the number of pieces on the board shrinks,
	we believe the value of piece differential will increase.
	Moreover, we believe that with fewer pieces on the board we are more likely to snapshot the game at a quiescent state,
	and avoid the error of counting material imbalance when it implicitly does not exist.
\item Similarly passed pawns become more valuable throughout the game. A passer on ranks 2 3 or 4 is a liability, with negative value,
	through much of the game. At these ranks, a passed pawn seems more like a target than a threat.
\item Generally the value of the (non passer) pawn is increasing faster than that of the major pieces, thus the relative value
	of the queen seems to decrease throughout the game.
\item Bishops are slightly more valuable than knights throughout the course of a game.
\item The major piece values are roughly consistent with the classical 3, 3, 5, 9 valuation.
\end{compactenum}

\begin{table}[!htbp] \centering 
  \caption{The estimated coefficients for standard chess are shown at different snapshots. The first column shows coefficients for the smaller formulation, repeating the middle column of \tabref{standard_simex_I_res_table}. The second column shows the coefficients for the expanded formulation, including passed pawns. The third column has results for games snapshotted in the first third of each game; the fourth column has snapshots in the middle third of each game; the fifth column has snapshots in the last third of each game. All results use the SIMEX + calibration procedure. } 
  \label{tab:standard_expand_tab_I} 
\small 
\begin{tabular}{@{\extracolsep{5pt}}lccccc} 
\\[-1.8ex]\hline 
\hline \\[-1.8ex] 
 & \multicolumn{5}{c}{\textit{Dependent variable:}} \\ 
\cline{2-6} 
\\[-1.8ex] & \multicolumn{5}{c}{Game Outcome} \\ 
 & \multicolumn{2}{c}{Random} & T1 & T2 & T3 \\ 
\\[-1.8ex] & (1) & (2) & (3) & (4) & (5)\\ 
\hline \\[-1.8ex] 
 $\Delta$ Glicko 2 & 0.94 & 0.94 & 1.02 & 0.96 & 0.81 \\ 
  White Advantage & 14.40 & 13.80 & 12.90 & 13.80 & 14.80 \\ 
  Tempo & 3.20 & 3.27 & $-$0.59 & 4.05 & 9.10 \\ 
  $\Delta$ Pawn & 67.60 & 56.10 & 19.90 & 49.50 & 69.40 \\ 
  $\Delta$ Knight & 195.00 & 195.00 & 113.00 & 172.00 & 217.00 \\ 
  $\Delta$ Bishop & 214.00 & 214.00 & 126.00 & 193.00 & 237.00 \\ 
  $\Delta$ Rook & 311.00 & 316.00 & 179.00 & 276.00 & 349.00 \\ 
  $\Delta$ Queen & 650.00 & 658.00 & 363.00 & 570.00 & 720.00 \\ 
  $\Delta$ Passed Pawn 2,3,4 &  & 7.61 & 14.20 & 6.14 & $-$0.04 \\ 
  $\Delta$ Passed Pawn 5 &  & 47.50 & 25.20 & 31.80 & 49.00 \\ 
  $\Delta$ Passed Pawn 6 &  & 93.70 & 9.77 & 42.40 & 105.00 \\ 
  $\Delta$ Passed Pawn 7 &  & 140.00 & $-$4.89 & 47.60 & 156.00 \\ 
 \hline \\[-1.8ex] 
num observations & 3,008,797 & 3,008,797 & 3,284,114 & 2,898,104 & 2,895,268 \\ 
\hline 
\hline \\[-1.8ex] 
\end{tabular} 
\end{table} 
\begin{sidewaystable}[!htbp] \centering 
  \caption{The estimated coefficients for standard chess are shown for different ranges of ply. All results use the SIMEX + calibration procedure. } 
  \label{tab:standard_ranged} 
\small 
\begin{tabular}{@{\extracolsep{5pt}}lccccccccc} 
\\[-1.8ex]\hline 
\hline \\[-1.8ex] 
 & \multicolumn{9}{c}{\textit{Dependent variable:}} \\ 
\cline{2-10} 
\\[-1.8ex] & \multicolumn{9}{c}{Game Outcome} \\ 
 & 10 - 19.5 & 20 - 29.5 & 30 - 39.5 & 40 - 49.5 & 50 - 59.5 & 60 - 69.5 & 70 - 79.5 & 80 - 89.5 & 90 - 99.5 \\ 
\\[-1.8ex] & (1) & (2) & (3) & (4) & (5) & (6) & (7) & (8) & (9)\\ 
\hline \\[-1.8ex] 
 $\Delta$ Glicko 2 & 1.04 & 0.98 & 0.92 & 0.85 & 0.77 & 0.71 & 0.65 & 0.58 & 0.54 \\ 
  White Advantage & 17.80 & 15.10 & 11.00 & 5.99 & 2.55 & $-$1.44 & $-$1.94 & $-$4.60 & $-$3.39 \\ 
  Tempo & 2.23 & 4.99 & 4.94 & 6.89 & 9.92 & 4.81 & 2.69 & 14.60 & $-$8.57 \\ 
  $\Delta$ Pawn & 22.90 & 44.10 & 52.50 & 61.80 & 69.40 & 77.60 & 85.50 & 91.60 & 91.90 \\ 
  $\Delta$ Knight & 97.00 & 147.00 & 170.00 & 192.00 & 217.00 & 242.00 & 261.00 & 264.00 & 262.00 \\ 
  $\Delta$ Bishop & 112.00 & 167.00 & 191.00 & 211.00 & 234.00 & 254.00 & 270.00 & 273.00 & 265.00 \\ 
  $\Delta$ Rook & 174.00 & 209.00 & 256.00 & 297.00 & 339.00 & 376.00 & 409.00 & 422.00 & 429.00 \\ 
  $\Delta$ Queen & 379.00 & 541.00 & 590.00 & 640.00 & 700.00 & 755.00 & 800.00 & 795.00 & 778.00 \\ 
  $\Delta$ Passed Pawn 2,3,4 & $-$33.10 & $-$15.80 & $-$10.30 & $-$12.00 & $-$8.93 & $-$6.16 & $-$3.86 & $-$1.21 & 6.14 \\ 
  $\Delta$ Passed Pawn 5 & 39.30 & 18.90 & 21.20 & 22.70 & 25.80 & 32.20 & 41.10 & 53.20 & 57.90 \\ 
  $\Delta$ Passed Pawn 6 & 31.80 & 60.30 & 61.60 & 69.70 & 71.00 & 80.10 & 86.20 & 99.90 & 111.00 \\ 
  $\Delta$ Passed Pawn 7 & 9.03 & 28.70 & 54.70 & 71.90 & 93.00 & 122.00 & 143.00 & 166.00 & 181.00 \\ 
 \hline \\[-1.8ex] 
num observations & 1,934,001 & 1,605,208 & 1,352,159 & 1,092,050 & 832,876 & 606,390 & 424,947 & 292,568 & 201,700 \\ 
mean ply & 14.3 & 24.3 & 34.3 & 44.3 & 54.2 & 64.2 & 74.2 & 84.2 & 94.1 \\ 
\hline 
\hline \\[-1.8ex] 
\end{tabular} 
\end{sidewaystable} 
\begin{knitrout}\small
\definecolor{shadecolor}{rgb}{0.969, 0.969, 0.969}\color{fgcolor}\begin{figure}[h]
\includegraphics[width=0.975\textwidth,height=1.024\textwidth]{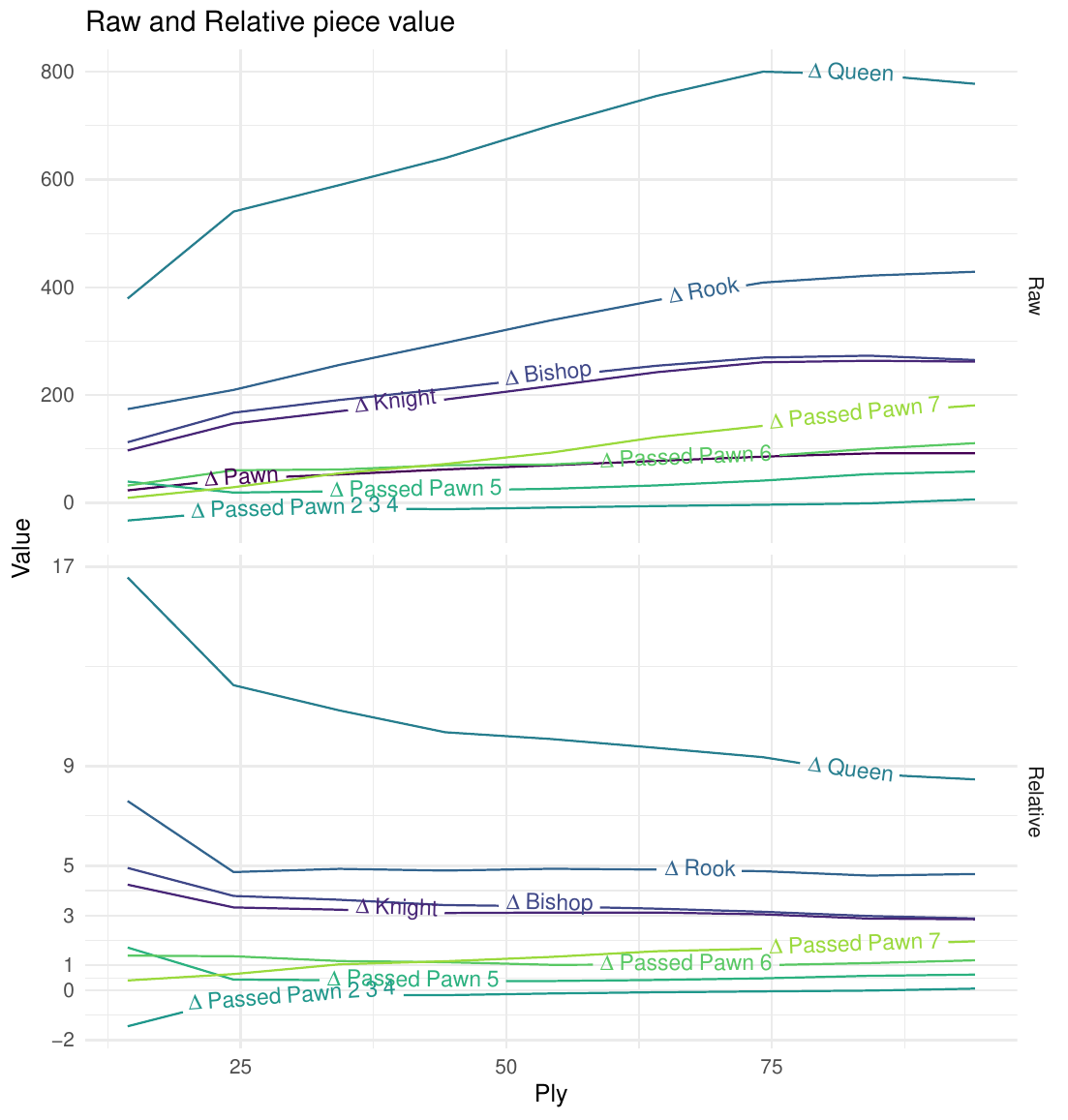} \caption[The value of piece imbalance for standard chess is plotted versus mean ply for the ranged estimations]{The value of piece imbalance for standard chess is plotted versus mean ply for the ranged estimations. The top facet is in raw units, while the bottom facet is relative to the absolute value of a pawn at that time. }\label{fig:standard_ranged_fit_plot}
\end{figure}

\end{knitrout}

Finally, we peel off the valuations from the earliest set of ply and
use those to quote values for equalizing games when one player is much stronger than another.
Rounding to 25 points, the
value of a pawn is approximately 25;
a knight is 100, as is a bishop;
a rook is worth 175;
and a queen is worth 375 Glicko-2 (or Elo) points.
By eyeball, however, it seems that if one were to extrapolate back from around 15 plies to 0 plies,
these equalizer values would be somewhat different. We leave this exercise to future research.

We note that using this system to even the odds of a game is subject to any positional imbalance caused.
For example, removing a pawn from the starting position may increase mobility for some of the major pieces which
would further affect the balance of the game.
It is probably better to not remove pawns and only use this system to remove major pieces when there is a 
large imbalance in player skill.

\clearpage

\subsection{Chess 960}

Before proceeding to consider Chess 960, we hazard some predictions.
Since the rules are the same as in standard chess, 
we expect that pieces will have the same value in the endgame in the two games.
Because players cannot rely on memorized openings, and some starting positions are awkward
given how the pieces move, we predict that more moves are be required in Chess 960
to reach the middlegame and endgame. 
Inasmuch as the ``luck'' of the opening position favors one player or another,
the value of pieces will be further depressed while the (unmeasured here) positional value will be
increased.
As such, we believe that the valuations in Chess 960 will resemble 
those in standard chess, but delayed and potentially muted.

Our dataset consists of Chess 960 games played on Lichess between August 2013 and May 2025.
We perform the same filters as we did on standard chess: 
we require both players to have at least 50 games in the database in the last six months,
filter on normal ending condition,
select games where each player has at least 5 minutes,
and require both players to have ratings of at least 1200.
This is a total of 
2,383,839 games.
From this set of games we perform the calibration procedure on Glicko-2 scores alone, as described above,
to estimate the $\sigma_0$, the amount of noise in the Glicko-2 ratings.
This we estimate as $\sigma_0=34$.
It is not clear why this is so much smaller than that for standard chess;
perhaps the code to compute ratings changed over time, or the wider time window used here provides
more stable estimates. 

As in the study of standard chess, we then select games where the
total plies is between 10 and 150 inclusive,
which leaves us with
2,329,586 games.
We then remove games with unusual recorded piece imbalance, using the same rules as in standard chess.
This results in a set of 2,329,508 games.
The median ending ply of games in this set is 60.

We then filter on no-recaptures in the next move, leaving
1,758,863 games, and perform the SIMEX and calibration procedure, resulting in the 
coefficients in \tabref{chess960_testing_t1}.
We also compute the coefficients on the random snapshots on the first third, middle third, and last third of the
matches, which are also given in that table.

We note that if we consider the regression uniformly selected over all plies, 
and divide the estimated piece values by the estimated pawn values, we
find that 
a knight is worth 2.7 pawns,
a bishop is worth 3 pawns,
a rook is worth 4.2 pawns,
and 
a queen is worth 8.7 pawns.
We plot these in \figref{standard_centipawn_plot_I}, along with the fit values from standard chess.
In comparison, the values, relative to a pawn, are slightly depressed in Chess 960.

\begin{table}[!htbp] \centering 
  \caption{The estimated coefficients for Chess 960 are shown at different snapshots. The first column shows coefficients for the uniformly random snapshot. The second column has results for games snapshotted in the first third of each game; the third column has snapshots in the middle third of each game; the fourth column has snapshots in the last third of each game. All results use the SIMEX + calibration procedure. } 
  \label{tab:chess960_testing_t1} 
\small 
\begin{tabular}{@{\extracolsep{5pt}}lcccc} 
\\[-1.8ex]\hline 
\hline \\[-1.8ex] 
 & \multicolumn{4}{c}{\textit{Dependent variable:}} \\ 
\cline{2-5} 
\\[-1.8ex] & \multicolumn{4}{c}{Game Outcome} \\ 
 & RR & T1 & T2 & T3 \\ 
\\[-1.8ex] & (1) & (2) & (3) & (4)\\ 
\hline \\[-1.8ex] 
  Glicko 2 & 0.96 & 1.01 & 0.97 & 0.86 \\ 
  White Advantage & 13.40 & 14.00 & 13.70 & 12.60 \\ 
  Tempo & 2.45 & $-$0.32 & 3.97 & 7.64 \\ 
   Pawn & 71.20 & 43.00 & 62.30 & 83.30 \\ 
   Knight & 222.00 & 145.00 & 200.00 & 243.00 \\ 
   Bishop & 245.00 & 169.00 & 224.00 & 264.00 \\ 
   Rook & 341.00 & 216.00 & 307.00 & 380.00 \\ 
   Queen & 713.00 & 456.00 & 641.00 & 784.00 \\ 
   Passed Pawn 2,3,4 & 2.93 & 0.18 & 0.31 & $-$3.09 \\ 
   Passed Pawn 5 & 37.60 & 11.40 & 26.00 & 41.10 \\ 
   Passed Pawn 6 & 88.20 & 15.20 & 42.90 & 101.00 \\ 
   Passed Pawn 7 & 140.00 & 30.50 & 56.00 & 159.00 \\ 
 \hline \\[-1.8ex] 
num observations & 1,758,863 & 1,948,197 & 1,692,683 & 1,663,135 \\ 
\hline 
\hline \\[-1.8ex] 
\end{tabular} 
\end{table} 

\begin{knitrout}\small
\definecolor{shadecolor}{rgb}{0.969, 0.969, 0.969}\color{fgcolor}\begin{figure}[h]
\includegraphics[width=0.975\textwidth,height=0.691\textwidth]{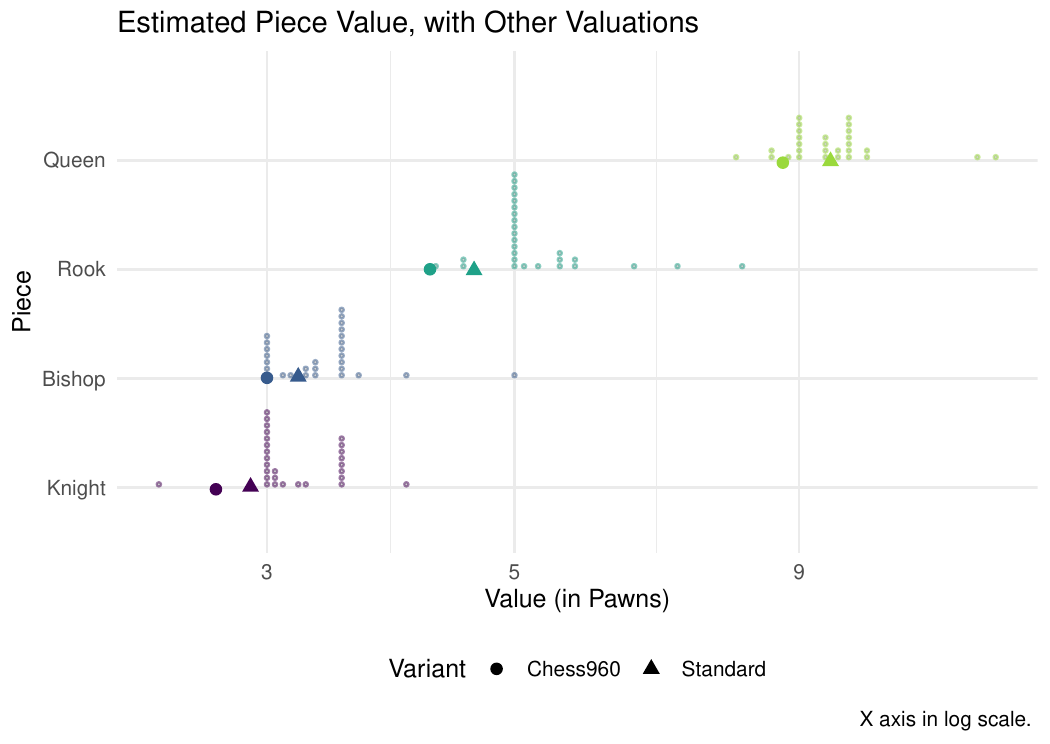} \caption[The calibrated estimated piece values, relative to a pawn, for standard chess and Chess 960 are plotted, along with the other valuation systems listed in \protect\tabref{value_systems}]{The calibrated estimated piece values, relative to a pawn, for standard chess and Chess 960 are plotted, along with the other valuation systems listed in \protect\tabref{value_systems}.}\label{fig:standard_centipawn_plot_I}
\end{figure}

\end{knitrout}

We also group games by ranges of ply, as we did in standard chess: looking among the four snapshots of a given game for one in the
given ply range, and which satisfies the no-capture condition.
We then perform the SIMEX and calibration procedure, and tabulate the results in \tabref{chess960_ranged}.
The values of the pieces are plotted in \figref{chess960_ranged_fit_plot} versus the mean ply of each slice of data,
both in raw values and relative to a pawn.
We see the same general shape and scale as in standard chess. 

We plot the by-ply-range estimates for Chess 960 alongside those for standard chess in \figref{chess960_and_standard_ranged_fit_plot}.
Contrary to our predictions, we see that piece values for Chess 960 are slightly higher than those for standard chess across all ply.
One could fit a just-so story consistent with these results: perhaps players of Chess 960 have less reliance on memorized positions,
and piece value is larger.  An alternative hypothesis is that the differences we see here are largely due to the differences 
in the estimated noise in ratings, $\sigma_0$. Further investigation is warranted.

\begin{sidewaystable}[!htbp] \centering 
  \caption{The estimated coefficients for Chess 960 are shown for different ranges of ply. All results use the SIMEX + calibration procedure. } 
  \label{tab:chess960_ranged} 
\small 
\begin{tabular}{@{\extracolsep{5pt}}lccccccccc} 
\\[-1.8ex]\hline 
\hline \\[-1.8ex] 
 & \multicolumn{9}{c}{\textit{Dependent variable:}} \\ 
\cline{2-10} 
\\[-1.8ex] & \multicolumn{9}{c}{Game Outcome} \\ 
 & 10 - 19.5 & 20 - 29.5 & 30 - 39.5 & 40 - 49.5 & 50 - 59.5 & 60 - 69.5 & 70 - 79.5 & 80 - 89.5 & 90 - 99.5 \\ 
\\[-1.8ex] & (1) & (2) & (3) & (4) & (5) & (6) & (7) & (8) & (9)\\ 
\hline \\[-1.8ex] 
  Glicko 2 & 1.03 & 1.00 & 0.95 & 0.88 & 0.81 & 0.74 & 0.67 & 0.58 & 0.52 \\ 
  White Advantage & 17.30 & 15.50 & 12.80 & 7.60 & 5.54 & 2.65 & 2.31 & $-$0.95 & $-$1.38 \\ 
  Tempo & 1.51 & 3.78 & 4.99 & 5.98 & 8.50 & 3.95 & 1.58 & 13.60 & $-$10.50 \\ 
   Pawn & 46.80 & 58.50 & 66.70 & 74.30 & 81.20 & 86.40 & 91.80 & 92.90 & 93.40 \\ 
   Knight & 164.00 & 181.00 & 196.00 & 217.00 & 239.00 & 263.00 & 279.00 & 287.00 & 282.00 \\ 
   Bishop & 196.00 & 208.00 & 221.00 & 238.00 & 255.00 & 271.00 & 285.00 & 289.00 & 281.00 \\ 
   Rook & 246.00 & 266.00 & 297.00 & 333.00 & 368.00 & 403.00 & 432.00 & 454.00 & 451.00 \\ 
   Queen & 559.00 & 600.00 & 652.00 & 713.00 & 761.00 & 813.00 & 846.00 & 842.00 & 811.00 \\ 
   Passed Pawn 2,3,4 & $-$11.40 & $-$10.30 & $-$10.80 & $-$11.90 & $-$11.10 & $-$6.26 & $-$1.72 & 3.83 & 12.20 \\ 
   Passed Pawn 5 & 5.62 & 14.70 & 22.70 & 22.70 & 26.10 & 31.20 & 40.20 & 54.30 & 58.50 \\ 
   Passed Pawn 6 & 35.70 & 51.40 & 64.10 & 72.70 & 76.70 & 80.30 & 90.40 & 107.00 & 112.00 \\ 
   Passed Pawn 7 & 52.60 & 62.20 & 50.60 & 83.50 & 110.00 & 125.00 & 158.00 & 177.00 & 194.00 \\ 
 \hline \\[-1.8ex] 
num observations & 1,164,366 & 955,905 & 782,824 & 617,823 & 456,474 & 324,492 & 222,404 & 150,834 & 101,900 \\ 
mean ply & 14.3 & 24.3 & 34.3 & 44.2 & 54.2 & 64.1 & 74.2 & 84.2 & 94.1 \\ 
\hline 
\hline \\[-1.8ex] 
\end{tabular} 
\end{sidewaystable} 
\begin{knitrout}\small
\definecolor{shadecolor}{rgb}{0.969, 0.969, 0.969}\color{fgcolor}\begin{figure}[h]
\includegraphics[width=0.975\textwidth,height=1.024\textwidth]{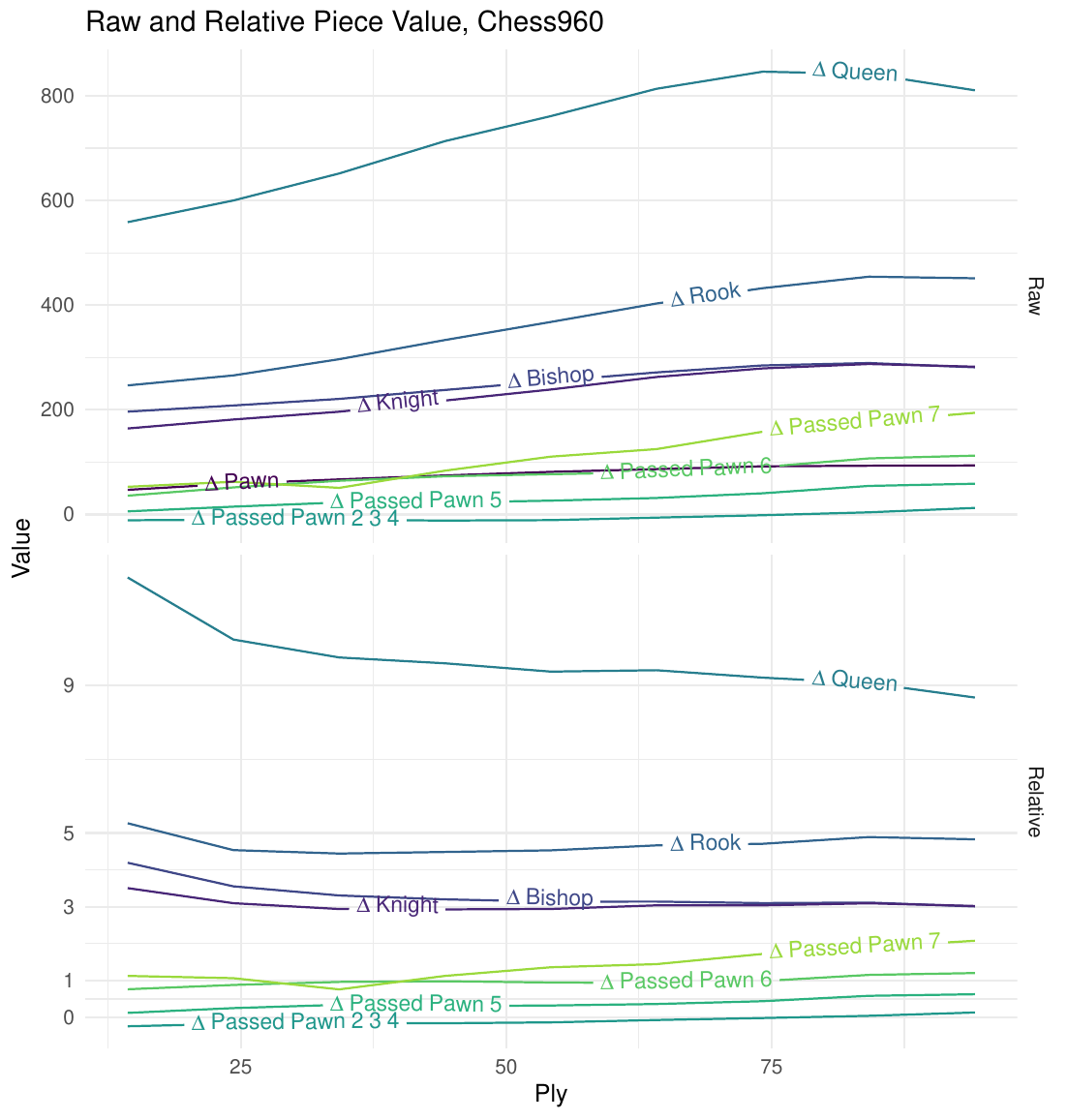} \caption[The value of piece imbalance for Chess 960 is plotted versus mean ply for the ranged estimations]{The value of piece imbalance for Chess 960 is plotted versus mean ply for the ranged estimations. The top facet is in raw units, while the bottom facet is relative to the absolute value of a pawn at that time. }\label{fig:chess960_ranged_fit_plot}
\end{figure}

\end{knitrout}
\begin{knitrout}\small
\definecolor{shadecolor}{rgb}{0.969, 0.969, 0.969}\color{fgcolor}\begin{figure}[h]
\includegraphics[width=0.975\textwidth,height=1.024\textwidth]{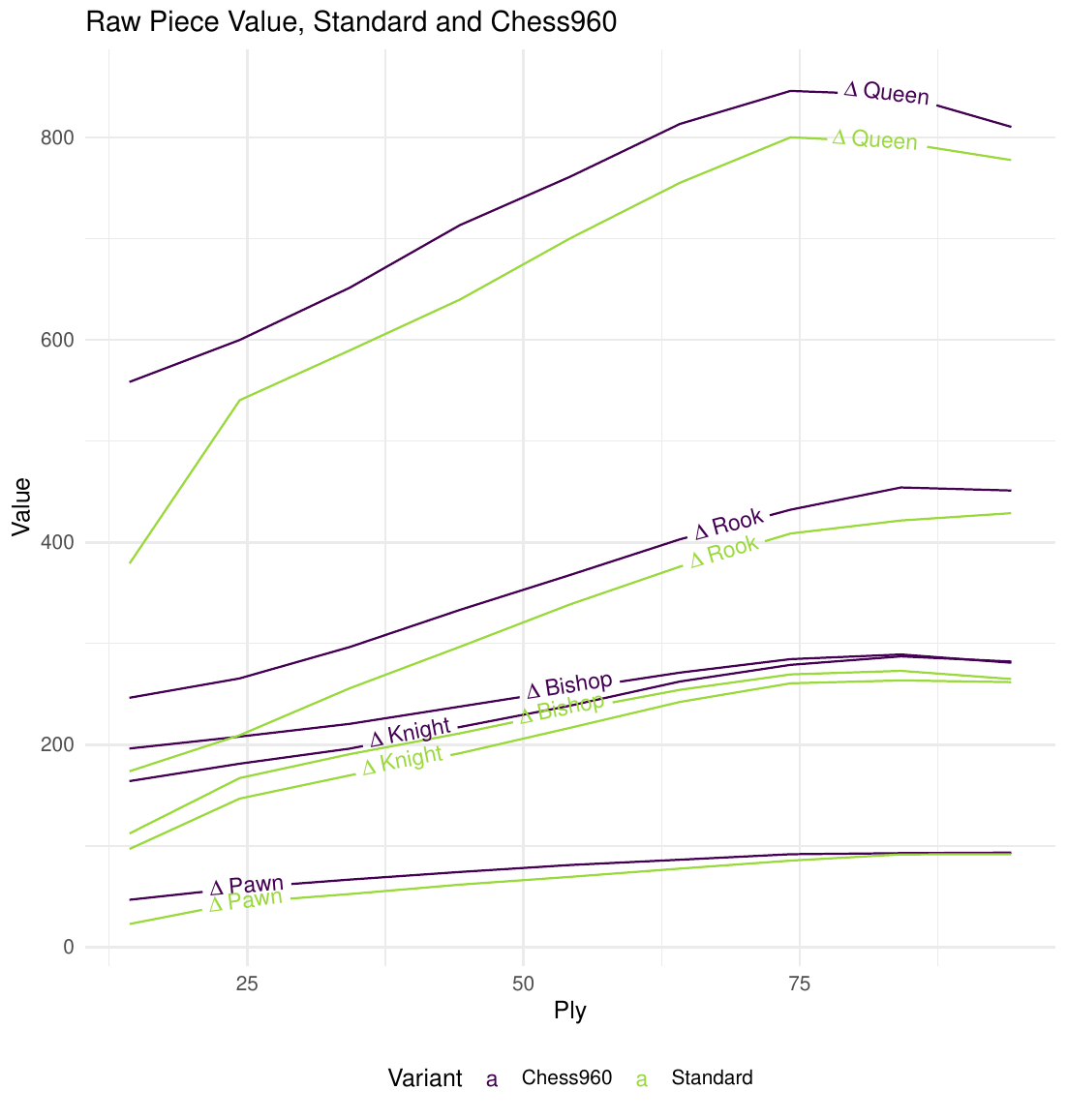} \caption[The values of piece imbalance for Chess 960 and standard chess are plotted versus mean ply for the ranged estimations]{The values of piece imbalance for Chess 960 and standard chess are plotted versus mean ply for the ranged estimations. }\label{fig:chess960_and_standard_ranged_fit_plot}
\end{figure}

\end{knitrout}

Finally, we peel off the valuations from the earliest set of ply and
use those to quote values for equalizing games when one player is much stronger than another.
Rounding to 25 points, the
value of a pawn is approximately 50;
a knight is 175;
a bishop is 200;
a rook is worth 250;
and a queen is worth 550 Glicko-2 (or Elo) points.
Again, these should probably be extrapolated back to zero ply.

\clearpage

\subsection{Atomic chess}

Our prediction for Atomic chess is that pawns and passed pawns will be relatively stronger than in standard chess,
as preventing promotion is more difficult, and pawns are more resistant to the explosions of capture.
Additionally, because a piece can remove multiple opposing pieces in one capture, piece imbalance is less predictive
than positional considerations.
Because the king cannot capture, less coordination is required to give checkmate, so a small piece imbalance can lead to victory;
this should also depress piece value.

Our dataset consists of Atomic chess games played on Lichess between January 2015 and May 2025.
We perform the same filters as we did on standard chess: 
we require both players to have at least 50 games in the database in the last six months,
filter on normal ending condition,
select games where each player has at least 5 minutes,
and require both players to have ratings of at least 1200.
This is a total of 
1,981,918 games.
From this set of games we perform the calibration procedure on Glicko-2 scores alone, as described above,
to estimate the $\sigma_0$, the amount of noise in the Glicko-2 ratings.
This we estimate as $\sigma_0=33$.
Again, it is not clear why this is smaller than the value estimated for standard chess.

As in the study of standard chess, we then select games where the
total plies is between 10 and 150 inclusive,
which leaves us with
1,759,702 games.
We then remove games with unusual recorded piece imbalance, using the same rules as in standard chess.
This results in a set of 1,759,376 games.
The median ending ply of games in this set is 30, which is considerably shorter than
in our analysis of standard chess and Chess 960.

We then filter on no-recaptures in the next move, leaving
1,510,059 games, and perform the SIMEX and calibration procedure, resulting in the 
coefficients in \tabref{atomic_testing_t1}.
We also compute the coefficients on the random snapshots on the first third, middle third, and last third of the
matches, which are also given in that table.

We note that if we consider the regression uniformly selected over all plies, 
and divide the estimated piece values by the estimated pawn values, we
find that 
a knight is worth 1.9 pawns,
a bishop is worth 2.5 pawns,
a rook is worth 4.9 pawns,
and 
a queen is worth 10 pawns.

We also group games by ranges of ply, as we did in standard chess: looking among the four snapshots of a given game for one in the
given ply range, and which satisfies the no-capture condition.
We then perform the SIMEX and calibration procedure, and tabulate the results in \tabref{atomic_ranged}.
The values of the pieces are plotted in \figref{atomic_ranged_fit_plot} versus the mean ply of each slice of data,
both in raw values and relative to the value of a pawn.
We plot the by ply range estimates for Atomic chess and standard chess in \figref{atomic_and_standard_ranged_fit_plot}.
We note the following:
\begin{compactenum}
\item Overall the value of pieces is much smaller in Atomic than in standard chess and Chess 960.
	We say that Atomic is a more \emph{positional} game than standard chess and Chess 960.
	Again we note that sometimes a major piece is pinned in their back rank preventing a pawn from promoting,
	so pieces on the board are sometimes of limited value in securing a win.
\item Despite the smaller absolute value of the major pieces, their relative values are similar to those in standard chess, with an average valuation closer to 2, 2.5, 5, 10.
\item As in standard chess, bishops are more valuable than knights.
\item The value of pieces seems to peak at around 60 ply. After this point, piece value decreases.
	This is in contrast to standard chess, where piece values appear to grow through 90 ply.
  The value of passed pawns also declines after this ply.
\item White's first player advantage is greater in Atomic chess than in standard, by a factor of around 3.
\end{compactenum}

\begin{table}[!htbp] \centering 
  \caption{The estimated coefficients for Atomic chess are shown at different snapshots. The first column shows coefficients for the uniformly random snapshot. The second column has results for games snapshotted in the first third of each game; the third column has snapshots in the middle third of each game; the fourth column has snapshots in the last third of each game. All results use the SIMEX + calibration procedure. } 
  \label{tab:atomic_testing_t1} 
\small 
\begin{tabular}{@{\extracolsep{5pt}}lcccc} 
\\[-1.8ex]\hline 
\hline \\[-1.8ex] 
 & \multicolumn{4}{c}{\textit{Dependent variable:}} \\ 
\cline{2-5} 
\\[-1.8ex] & \multicolumn{4}{c}{Game Outcome} \\ 
 & RR & T1 & T2 & T3 \\ 
\\[-1.8ex] & (1) & (2) & (3) & (4)\\ 
\hline \\[-1.8ex] 
  Glicko 2 & 0.91 & 0.97 & 0.91 & 0.85 \\ 
  White Advantage & 44.10 & 53.10 & 41.20 & 38.60 \\ 
  Tempo & 0.08 & $-$2.21 & $-$2.17 & 4.47 \\ 
   Pawn & 21.80 & 38.30 & 26.80 & 16.50 \\ 
   Knight & 41.40 & 65.10 & 49.60 & 28.60 \\ 
   Bishop & 55.00 & 60.60 & 58.10 & 53.40 \\ 
   Rook & 106.00 & 86.80 & 103.00 & 110.00 \\ 
   Queen & 222.00 & 121.00 & 188.00 & 260.00 \\ 
   Passed Pawn 2,3,4 & 18.00 & 39.10 & 40.40 & 12.10 \\ 
   Passed Pawn 5 & 27.40 & 22.80 & 41.30 & 28.40 \\ 
   Passed Pawn 6 & 36.60 & 4.81 & 39.90 & 41.10 \\ 
   Passed Pawn 7 & 87.10 & 8.77 & 53.70 & 98.60 \\ 
 \hline \\[-1.8ex] 
num observations & 1,510,059 & 1,623,393 & 1,502,340 & 1,455,459 \\ 
\hline 
\hline \\[-1.8ex] 
\end{tabular} 
\end{table} 
\begin{sidewaystable}[!htbp] \centering 
  \caption{The estimated coefficients for Atomic chess are shown for different ranges of ply. All results use the SIMEX + calibration procedure. } 
  \label{tab:atomic_ranged} 
\small 
\begin{tabular}{@{\extracolsep{5pt}}lccccccccc} 
\\[-1.8ex]\hline 
\hline \\[-1.8ex] 
 & \multicolumn{9}{c}{\textit{Dependent variable:}} \\ 
\cline{2-10} 
\\[-1.8ex] & \multicolumn{9}{c}{Game Outcome} \\ 
 & 10 - 19.5 & 20 - 29.5 & 30 - 39.5 & 40 - 49.5 & 50 - 59.5 & 60 - 69.5 & 70 - 79.5 & 80 - 89.5 & 90 - 99.5 \\ 
\\[-1.8ex] & (1) & (2) & (3) & (4) & (5) & (6) & (7) & (8) & (9)\\ 
\hline \\[-1.8ex] 
  Glicko 2 & 0.93 & 0.79 & 0.70 & 0.63 & 0.55 & 0.47 & 0.42 & 0.41 & 0.34 \\ 
  White Advantage & 45.40 & 20.40 & 14.60 & 12.90 & 10.60 & 5.08 & 5.01 & 4.97 & 7.56 \\ 
  Tempo & $-$1.40 & 4.96 & 3.68 & 5.99 & 4.53 & $-$4.94 & $-$6.88 & 10.60 & $-$13.50 \\ 
   Pawn & 25.90 & 21.20 & 29.40 & 45.30 & 61.80 & 68.50 & 60.40 & 42.20 & 37.00 \\ 
   Knight & 24.10 & 48.30 & 70.20 & 94.40 & 123.00 & 135.00 & 120.00 & 82.80 & 69.60 \\ 
   Bishop & 32.60 & 59.80 & 87.80 & 114.00 & 136.00 & 144.00 & 127.00 & 90.80 & 65.60 \\ 
   Rook & 36.10 & 84.30 & 126.00 & 173.00 & 217.00 & 248.00 & 227.00 & 185.00 & 152.00 \\ 
   Queen & 153.00 & 243.00 & 276.00 & 314.00 & 375.00 & 402.00 & 339.00 & 286.00 & 257.00 \\ 
   Passed Pawn 2,3,4 & $-$34.20 & $-$4.41 & 3.76 & 8.02 & 9.25 & 4.84 & $-$0.02 & 4.88 & $-$1.72 \\ 
   Passed Pawn 5 & 1.36 & 7.59 & 25.70 & 36.60 & 44.70 & 41.80 & 24.60 & 14.80 & 9.23 \\ 
   Passed Pawn 6 & 62.50 & 49.80 & 47.50 & 47.70 & 55.60 & 57.70 & 40.60 & 26.20 & 14.20 \\ 
   Passed Pawn 7 & 284.00 & 132.00 & 96.10 & 97.30 & 110.00 & 119.00 & 105.00 & 79.90 & 66.40 \\ 
 \hline \\[-1.8ex] 
num observations & 1,182,700 & 698,071 & 413,599 & 249,971 & 151,913 & 91,826 & 52,786 & 29,715 & 17,114 \\ 
mean ply & 13.8 & 23.9 & 34 & 44 & 54 & 64 & 74 & 84 & 94 \\ 
\hline 
\hline \\[-1.8ex] 
\end{tabular} 
\end{sidewaystable} 
\begin{knitrout}\small
\definecolor{shadecolor}{rgb}{0.969, 0.969, 0.969}\color{fgcolor}\begin{figure}[h]
\includegraphics[width=0.975\textwidth,height=1.024\textwidth]{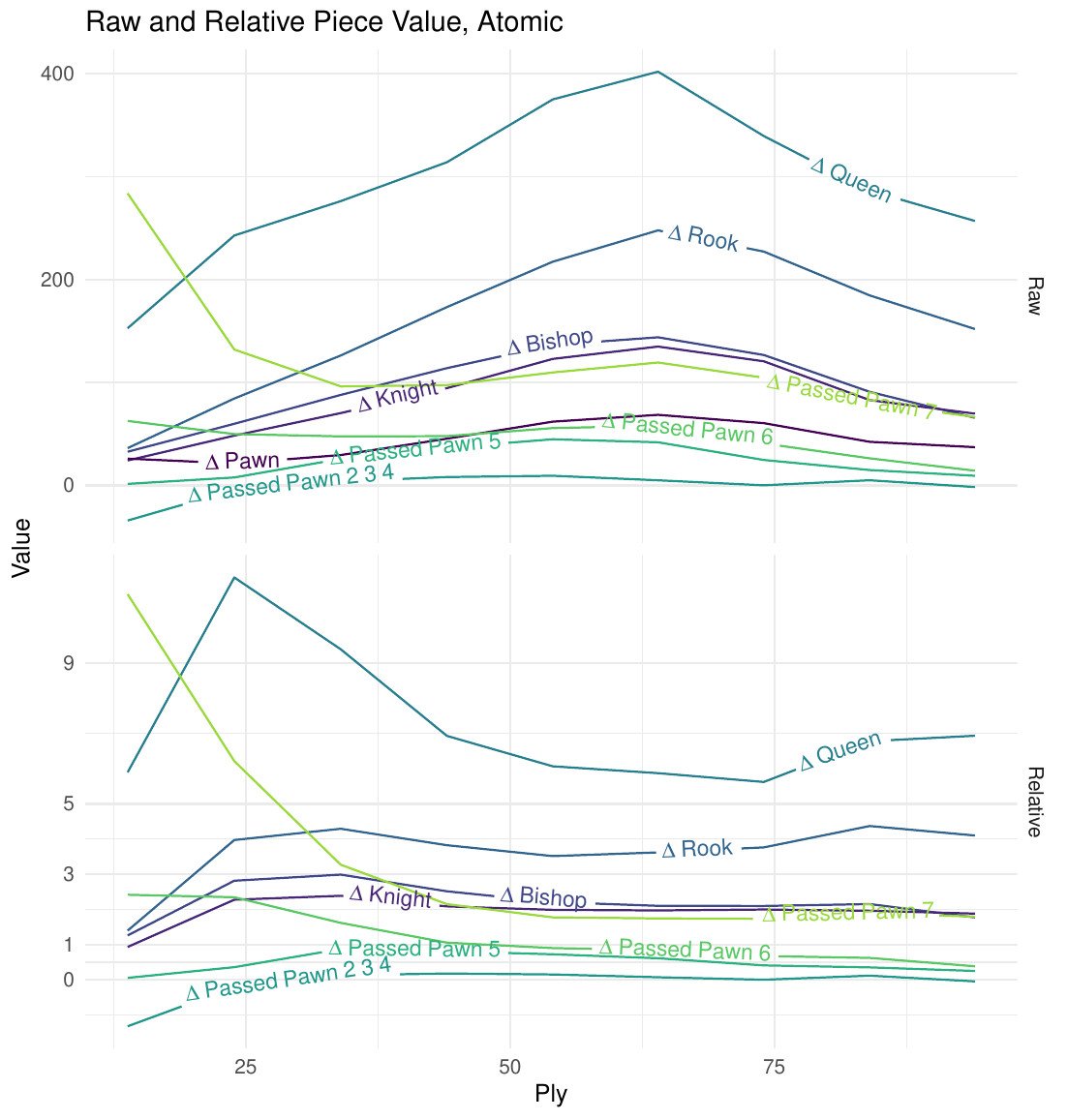} \caption[The value of piece imbalance for Atomic chess is plotted versus mean ply for the ranged estimations]{The value of piece imbalance for Atomic chess is plotted versus mean ply for the ranged estimations. The top facet is in raw units, while the bottom facet is relative to the absolute value of a pawn at that time. }\label{fig:atomic_ranged_fit_plot}
\end{figure}

\end{knitrout}
\begin{knitrout}\small
\definecolor{shadecolor}{rgb}{0.969, 0.969, 0.969}\color{fgcolor}\begin{figure}[h]
\includegraphics[width=0.975\textwidth,height=1.024\textwidth]{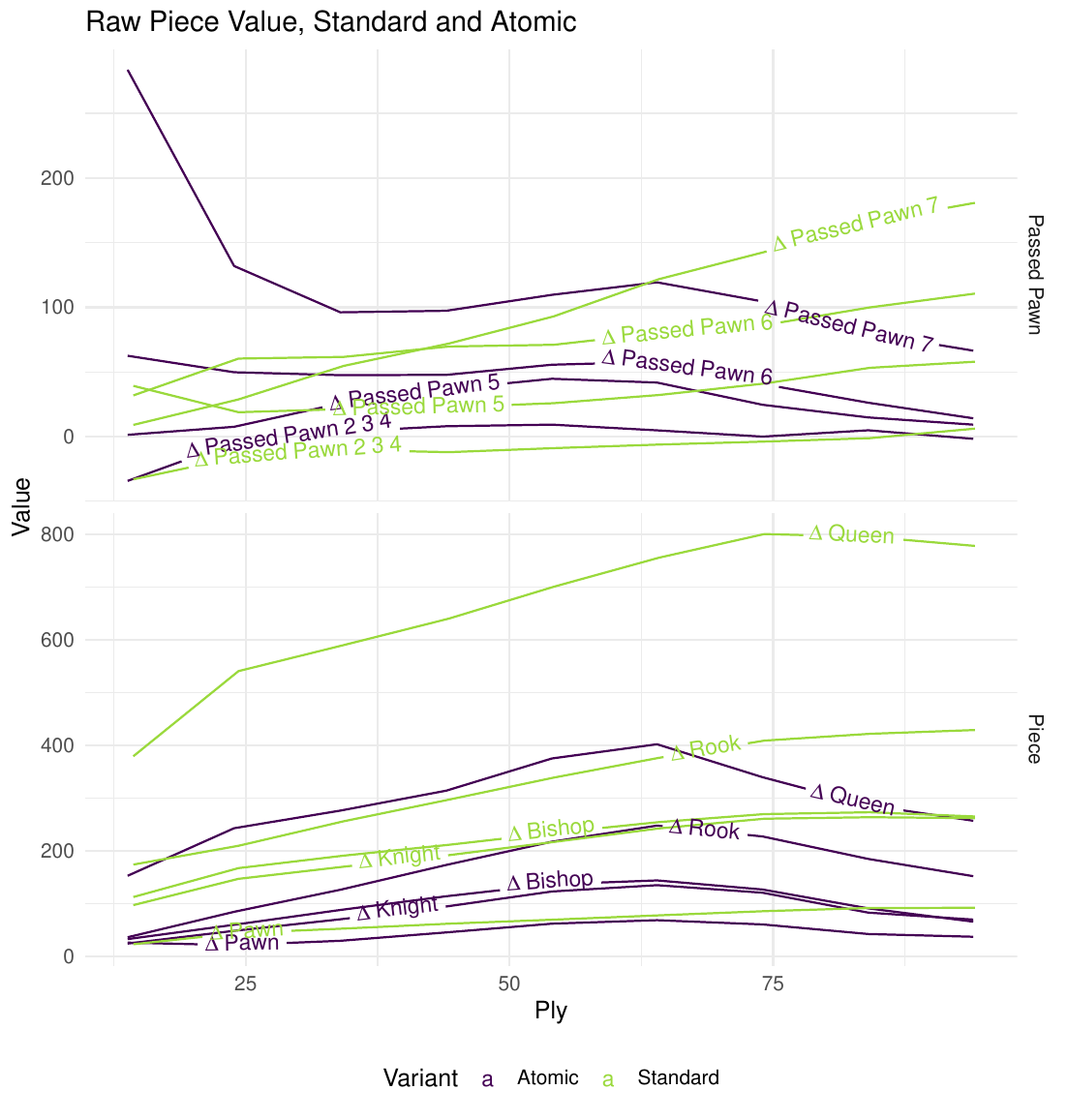} \caption[The values of piece imbalance for Atomic and standard chess are plotted versus mean ply for the ranged estimations]{The values of piece imbalance for Atomic and standard chess are plotted versus mean ply for the ranged estimations. }\label{fig:atomic_and_standard_ranged_fit_plot}
\end{figure}

\end{knitrout}

Finally, we peel off the valuations from the earliest set of ply and
use those to quote values for equalizing games when one player is much stronger than another.
Rounding to 25 points, the
pawn, knight, bishop and rook are all worth 25,
and a queen is worth 150 Glicko-2 (or Elo) points.
Again, these should probably be extrapolated back to zero ply.
More so than in standard chess or Chess 960,
equalization must be performed carefully, as a hole in the pawn defenses could
give potentially free a major piece or expose the king in such a way as to be
decisive. 
Equalization by material should be tested via computer play.

\clearpage

\subsection{Antichess}

We make some predictions regarding Antichess.
First ``piece value'' should be \emph{negative}, since the winning condition is tied to
losing all your pieces.
We predict that bishops will be fairly negative, especially in early game, as they are
often drawn into chains of captures.
We predict the king will have a relatively high value: the king has many possible moves, but
cannot easily be drawn into long range capture chains.
As noted above, the same piece imbalance but with different position is either winning for white or black. 
Moreover having tempo is often a deciding factor for a given position.
For this reason we believe piece values will be compressed compared to standard chess.
As white theoretically has a win from the opening, we expect that white's first player advantage will be significant.
\cite{watkins2017losing}

Our dataset consists of Antichess games played on Lichess between December 2014 and May 2025.
We perform the same filters as we did on standard chess: 
we require both players to have at least 50 games in the database in the last six months,
filter on normal ending condition,
select games where each player has at least 5 minutes,
and require both players to have ratings of at least 1200.
This is a total of 
3,440,433 games.
From this set of games we perform the calibration procedure on Glicko-2 scores alone, as described above,
to estimate the $\sigma_0$, the amount of noise in the Glicko-2 ratings.
This we estimate as $\sigma_0=67$.
It is not clear why this is higher than the value found for standard chess.

As in the study of standard chess, we then select games where the
total plies is between 10 and 150 inclusive,
which leaves us with
3,415,636 games.
We then remove games with unusual recorded piece imbalance, using the same rules as in standard chess.
This results in a set of 3,415,604 games.
The median ending ply of games in this set is 47, shorter than 
in our analysis of standard chess and Chess 960, but longer than atomic chess.

We then filter on no-recaptures in the next move, leaving
1,589,198 games, and perform the SIMEX and calibration procedure, resulting in the 
coefficients in \tabref{antichess_testing_t1}.
We also compute the coefficients on the random snapshots on the first third, middle third, and last third of the
matches, which are also given in that table.

We note that if we consider the regression uniformly selected over all plies, 
and divide the estimated piece values by the estimated pawn values (which is negative), 
we find that 
a knight is worth 0.77 pawns,
a bishop is worth 1.2 pawns,
a rook is worth 1 pawns,
a queen is worth 1.4 pawns, 
and 
a king is worth -0.041 pawns.
The negative relative value for the king here indicates that a king has slightly positive average value, though very close to zero.
The king takes negative value in the last third of games, but otherwise has slightly positive value.

We also group games by ranges of ply, as we did above, by looking among the four snapshots of a given game for one in the
given ply range, and which satisfies the no-capture condition.
We then perform the SIMEX and calibration procedure, and tabulate the results in \tabref{antichess_ranged}.
The values of the pieces are plotted in \figref{antichess_ranged_fit_plot} versus the mean ply of each slice of data,
both in raw values and relative to the value of a pawn.
We note the following:
\begin{compactenum}
\item White's first player advantage is actually fairly modest, smaller than in Atomic chess.
	This is despite the fact that Antichess is theoretically winning for white.
\item Most pieces have negative value, though the king appears to have positive value in the middlegame,
	and bishops and rooks have positive value in the endgame.
	Again this is conditional on no immediate capture in the next move, which would preclude
	snapshots where the rook or bishop are already in the midst of a capture chain.
	These findings are consistent with commonly played strategies.
\item Unlike in standard chess, and the other variants considered here, 
	the valuations of knights and bishops are very different from each other as a function of ply.
	As noted above, knights are a liability throughout the range of ply tested here, while
	bishops, on average, are valuable in an endgame. As with all the findings here, this is contingent
	on the position.
\item Pawns are a greater liability in the endgame. 
	Anecdotally, it is especially dangerous to hold only pawns in the endgame,
	as they have very constrained movement.
\item In general the absolute value of material imbalance is quite a bit smaller than in standard chess and
	the other variants we consider. Antichess is a positional game, more so than Atomic.
\end{compactenum}

\begin{table}[!htbp] \centering 
  \caption{The estimated coefficients for Antichess are shown at different snapshots. The first column shows coefficients for the uniformly random snapshot. The second column has results for games snapshotted in the first third of each game; the third column has snapshots in the middle third of each game; the fourth column has snapshots in the last third of each game. All results use the SIMEX + calibration procedure. } 
  \label{tab:antichess_testing_t1} 
\small 
\begin{tabular}{@{\extracolsep{5pt}}lcccc} 
\\[-1.8ex]\hline 
\hline \\[-1.8ex] 
 & \multicolumn{4}{c}{\textit{Dependent variable:}} \\ 
\cline{2-5} 
\\[-1.8ex] & \multicolumn{4}{c}{Game Outcome} \\ 
 & RR & T1 & T2 & T3 \\ 
\\[-1.8ex] & (1) & (2) & (3) & (4)\\ 
\hline \\[-1.8ex] 
  Glicko 2 & 0.86 & 0.99 & 0.89 & 0.68 \\ 
  White Advantage & 13.30 & 10.40 & 8.35 & 9.20 \\ 
  Tempo & $-$54.20 & $-$8.47 & $-$45.60 & $-$112.00 \\ 
   Pawn & $-$49.30 & $-$8.56 & $-$17.50 & $-$88.00 \\ 
   Knight & $-$37.80 & $-$2.05 & $-$9.78 & $-$70.30 \\ 
   Bishop & $-$58.40 & $-$26.90 & $-$47.00 & $-$71.50 \\ 
   Rook & $-$51.00 & $-$17.10 & $-$38.30 & $-$32.50 \\ 
   Queen & $-$66.80 & $-$21.40 & $-$71.50 & $-$140.00 \\ 
   King & 2.02 & 17.00 & 12.30 & $-$13.40 \\ 
 \hline \\[-1.8ex] 
num observations & 1,589,198 & 1,318,020 & 1,547,890 & 1,886,375 \\ 
\hline 
\hline \\[-1.8ex] 
\end{tabular} 
\end{table} 
\begin{sidewaystable}[!htbp] \centering 
  \caption{The estimated coefficients for Antichess are shown for different ranges of ply. All results use the SIMEX + calibration procedure. } 
  \label{tab:antichess_ranged} 
\small 
\begin{tabular}{@{\extracolsep{5pt}}lccccccccc} 
\\[-1.8ex]\hline 
\hline \\[-1.8ex] 
 & \multicolumn{9}{c}{\textit{Dependent variable:}} \\ 
\cline{2-10} 
\\[-1.8ex] & \multicolumn{9}{c}{Game Outcome} \\ 
 & 10 - 19.5 & 20 - 29.5 & 30 - 39.5 & 40 - 49.5 & 50 - 59.5 & 60 - 69.5 & 70 - 79.5 & 80 - 89.5 & 90 - 99.5 \\ 
\\[-1.8ex] & (1) & (2) & (3) & (4) & (5) & (6) & (7) & (8) & (9)\\ 
\hline \\[-1.8ex] 
  Glicko 2 & 1.01 & 0.87 & 0.73 & 0.64 & 0.60 & 0.56 & 0.51 & 0.44 & 0.37 \\ 
  White Advantage & 16.00 & 9.33 & 7.41 & 4.77 & 4.79 & 3.64 & 2.95 & 4.70 & 6.51 \\ 
  Tempo & $-$33.30 & $-$58.20 & $-$95.10 & $-$86.60 & $-$62.00 & $-$40.00 & $-$32.20 & $-$51.10 & $-$10.10 \\ 
   Pawn & $-$32.70 & $-$32.40 & $-$41.60 & $-$57.90 & $-$77.90 & $-$97.40 & $-$103.00 & $-$88.40 & $-$80.70 \\ 
   Knight & $-$38.10 & $-$34.70 & $-$31.00 & $-$31.60 & $-$44.30 & $-$82.20 & $-$121.00 & $-$120.00 & $-$101.00 \\ 
   Bishop & $-$67.90 & $-$70.10 & $-$65.70 & $-$45.80 & $-$11.60 & 10.10 & 14.00 & 19.80 & 20.80 \\ 
   Rook & $-$80.20 & $-$94.90 & $-$75.20 & $-$46.20 & $-$5.49 & 44.60 & 69.10 & 84.30 & 98.20 \\ 
   Queen & $-$51.50 & $-$73.10 & $-$89.60 & $-$74.20 & $-$77.00 & $-$44.70 & $-$47.70 & $-$50.80 & $-$57.10 \\ 
   King & $-$40.90 & $-$1.17 & 18.60 & 19.90 & 9.73 & $-$17.30 & $-$35.30 & $-$27.90 & $-$20.60 \\ 
 \hline \\[-1.8ex] 
num observations & 985,732 & 1,096,642 & 1,076,260 & 701,170 & 430,904 & 260,901 & 125,455 & 52,394 & 22,045 \\ 
mean ply & 14.7 & 24.5 & 34.2 & 44 & 54.1 & 63.9 & 73.8 & 83.8 & 93.8 \\ 
\hline 
\hline \\[-1.8ex] 
\end{tabular} 
\end{sidewaystable} 
\begin{knitrout}\small
\definecolor{shadecolor}{rgb}{0.969, 0.969, 0.969}\color{fgcolor}\begin{figure}[h]
\includegraphics[width=0.975\textwidth,height=1.024\textwidth]{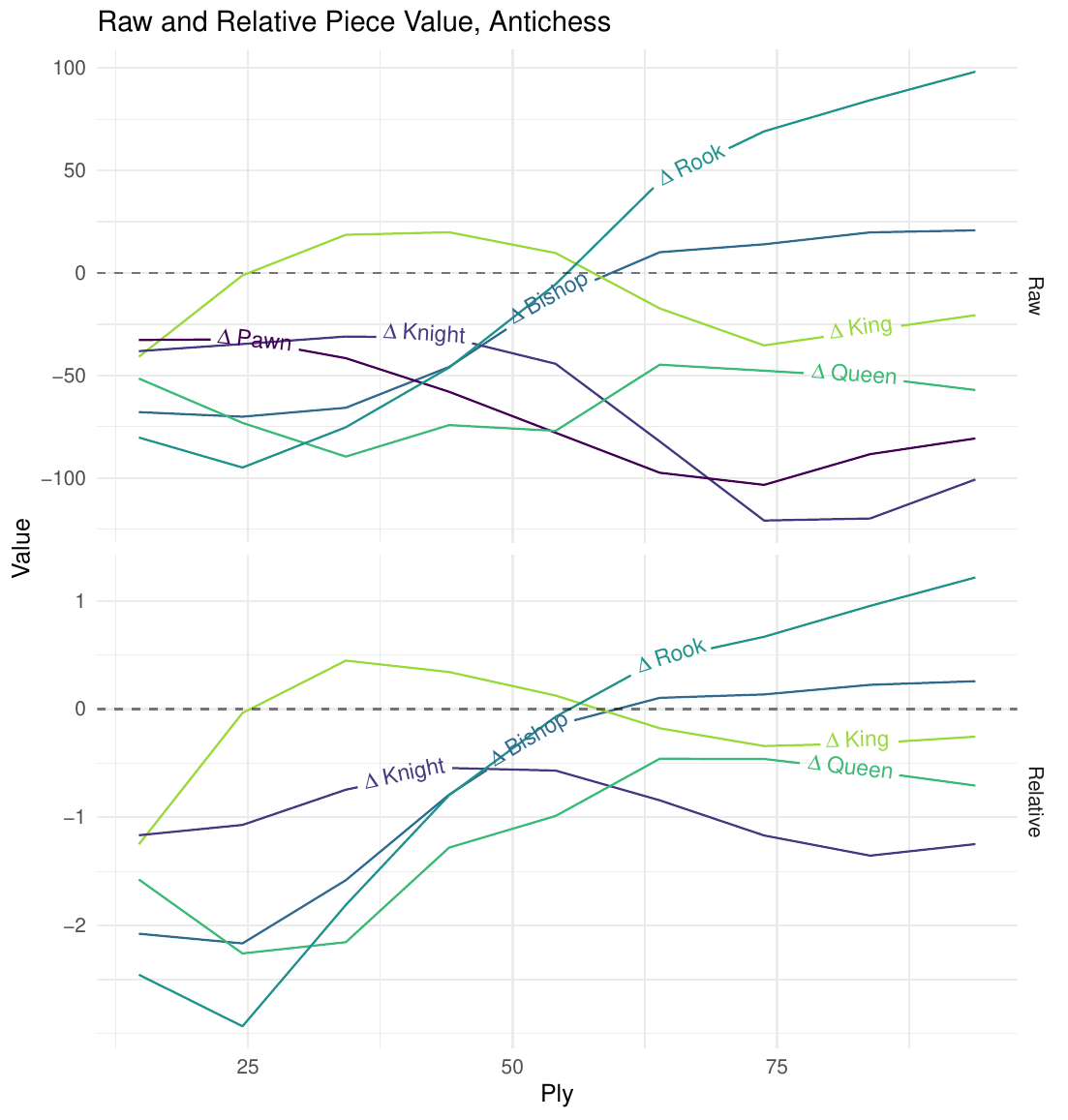} \caption[The value of piece imbalance for Antichess is plotted versus mean ply for the ranged estimations]{The value of piece imbalance for Antichess is plotted versus mean ply for the ranged estimations. The top facet is in raw units, while the bottom facet is relative to the absolute value of a pawn at that time. }\label{fig:antichess_ranged_fit_plot}
\end{figure}

\end{knitrout}

Finally, we peel off the valuations from the earliest set of ply and
use those to quote values for equalizing games when one player is much stronger than another.
Rounding to 25 points, the
value of a pawn is approximately -25;
a bishop and a rook are both worth -75;
and a knight, queen and a king are each worth -50
Glicko-2 (or Elo) points.
We note that using this system to even the odds of a game is subject to any positional imbalance caused.
For example, removing a pawn from the starting position may increase mobility for some of the major pieces which
would further affect the balance of the game.
It is probably better to not remove pawns to achieve balance and only use this system to remove major pieces when there is a 
large imbalance in player skill.

\clearpage

\subsection{Horde}

We hazard few predictions about Horde chess.
Anecdotally, it seems that black often trades away bishops and knights for fewer than three white pawns each,
that rooks are particularly dangerous, as white's outer file pawns can more easily be picked off.
We suspect that Horde is more positional than standard chess.

Our dataset consists of Horde games played on Lichess between December 2014 and May 2025.
We perform the same filters as we did on standard chess: 
we require both players to have at least 50 games in the database in the last six months,
filter on normal ending condition,
select games where each player has at least 5 minutes,
and require both players to have ratings of at least 1200.
This is a total of 
395,207 games.
From this set of games we perform the calibration procedure on Glicko-2 scores alone, as described above,
to estimate the $\sigma_0$, the amount of noise in the Glicko-2 ratings.
This we estimate as $\sigma_0=34$.

As in the study of standard chess, we then select games where the
total plies is between 10 and 150 inclusive,
which leaves us with
379,417 games.
We then remove games with unusual recorded piece imbalance, using the same rules as in standard chess.
This results in a set of 379,279 games.
The median ending ply of games in this set is 98, which is quite a bit longer than
in our analysis of standard chess and Chess 960, but not surprising given the number of pieces on the
board and the pace of a typical match.

We then filter on no-recaptures in the next move, leaving
227,798 games, and perform the SIMEX and calibration procedure, resulting in the 
coefficients in \tabref{horde_testing_t1}.
We also compute the coefficients on the random snapshots on the first third, middle third, and last third of the
matches, which are also given in that table.
Here we regress using a simplified specification, with only piece imbalance, omitting the imbalance
of passed pawns.
Black is unlikely to ever have a passed pawn, and white is likely to have a passed pawn blocked by another passed pawn, 
which would not be properly recorded in our calculations.

\begin{table}[!htbp] \centering 
  \caption{The estimated coefficients for Horde are shown at different snapshots. The first column shows coefficients for the uniformly random snapshot. The second column has results for games snapshotted in the first third of each game; the third column has snapshots in the middle third of each game; the fourth column has snapshots in the last third of each game. All results use the SIMEX + calibration procedure. } 
  \label{tab:horde_testing_t1} 
\small 
\begin{tabular}{@{\extracolsep{5pt}}lcccc} 
\\[-1.8ex]\hline 
\hline \\[-1.8ex] 
 & \multicolumn{4}{c}{\textit{Dependent variable:}} \\ 
\cline{2-5} 
\\[-1.8ex] & \multicolumn{4}{c}{Game Outcome} \\ 
 & RR & T1 & T2 & T3 \\ 
\\[-1.8ex] & (1) & (2) & (3) & (4)\\ 
\hline \\[-1.8ex] 
  Glicko 2 & 0.91 & 0.99 & 0.90 & 0.79 \\ 
  White Advantage & $-$113.00 & $-$1,200.00 & $-$585.00 & $-$128.00 \\ 
  Tempo & 15.50 & 5.05 & 16.50 & 29.30 \\ 
   Pawn & 59.90 & 86.20 & 73.30 & 69.30 \\ 
   Knight & 146.00 & 139.00 & 143.00 & 170.00 \\ 
   Bishop & 157.00 & 148.00 & 154.00 & 190.00 \\ 
   Rook & 219.00 & 127.00 & 161.00 & 253.00 \\ 
   Queen & 562.00 & 416.00 & 534.00 & 561.00 \\ 
 \hline \\[-1.8ex] 
num observations & 227,798 & 223,199 & 227,371 & 234,890 \\ 
\hline 
\hline \\[-1.8ex] 
\end{tabular} 
\end{table} 

We note that the ``White Advantage'' term is negative and has very large magnitude.
One should not conclude from this that white is at a disadvantage in Horde.
Rather this is due to the non-zero mean in the imbalanced covariates:
white is likely to have many more pawns, and far fewer major pieces than black.
If you use the first third regression estimates on the piece imbalance at the start
of a Horde game (white is up 28 pawns, down 2 pawns, knights and rooks, down 1 queen, and has the tempo), 
the estimated effect for white is -32, in Glicko-2 units.
This is rather modest.
For the ranged fits, described below, the estimated effect for white is -33, in Glicko-2 units.

We note that if we consider the regression uniformly selected over all plies, 
and divide the estimated piece values by the estimated pawn values (which is negative), 
we find that 
a knight is worth 2.4 pawns,
a bishop is worth 2.6 pawns,
a rook is worth 3.7 pawns,
and 
a queen is worth 9.4 pawns.

We also group games by ranges of ply, as we did above, by looking among the four snapshots of a given game for one in the
given ply range, and which satisfies the no-capture condition.
We then perform the SIMEX and calibration procedure, and tabulate the results in \tabref{horde_shorter_ranged}.
The values of the pieces are plotted in \figref{horde_shorter_ranged_fit_plot} versus the mean ply of each slice of data,
both in raw values and relative to the value of a pawn.
We note the following:
\begin{compactenum}
\item As in other variants, a bishop is worth slightly more than a knight.
\item Somewhat surprisingly the rook is worth about as much as a bishop or knight during the
	early phase of the game. Rooks are only more valuable than bishops or knights in the later ply.
\item The bishop, knight and rook have depressed value, compared to a pawn, while the queen is still highly valuable.
	The major pieces are valued roughly as 2.5, 2.5, 4, 9.
\item The value of a pawn decreases slightly over the plies tested. White tends to shed their pawns, while black must
	promote to win, so this is not surprising.
\item It is hard to interpret the changes in white's first player advantage over the course of a match.
	It is roughly decreasing in magnitude as white's pawn imbalance also decreases.
\end{compactenum}

\begin{sidewaystable}[!htbp] \centering 
  \caption{The estimated coefficients for Horde are shown for different ranges of ply. All results use the SIMEX + calibration procedure. } 
  \label{tab:horde_shorter_ranged} 
\small 
\begin{tabular}{@{\extracolsep{5pt}}lccccccccc} 
\\[-1.8ex]\hline 
\hline \\[-1.8ex] 
 & \multicolumn{9}{c}{\textit{Dependent variable:}} \\ 
\cline{2-10} 
\\[-1.8ex] & \multicolumn{9}{c}{Game Outcome} \\ 
 & 10 - 19.5 & 20 - 29.5 & 30 - 39.5 & 40 - 49.5 & 50 - 59.5 & 60 - 69.5 & 70 - 79.5 & 80 - 89.5 & 90 - 99.5 \\ 
\\[-1.8ex] & (1) & (2) & (3) & (4) & (5) & (6) & (7) & (8) & (9)\\ 
\hline \\[-1.8ex] 
  Glicko 2 & 1.04 & 0.96 & 0.90 & 0.85 & 0.78 & 0.74 & 0.71 & 0.64 & 0.55 \\ 
  White Advantage & $-$1,530.00 & $-$1,810.00 & $-$1,580.00 & $-$1,140.00 & $-$696.00 & $-$437.00 & $-$291.00 & $-$257.00 & $-$219.00 \\ 
  Tempo & 7.58 & 6.83 & 12.60 & 15.40 & 18.60 & 22.30 & 22.80 & 32.70 & 19.90 \\ 
   Pawn & 109.00 & 120.00 & 114.00 & 102.00 & 83.60 & 74.60 & 70.40 & 71.50 & 74.40 \\ 
   Knight & 178.00 & 167.00 & 159.00 & 148.00 & 127.00 & 118.00 & 112.00 & 121.00 & 134.00 \\ 
   Bishop & 180.00 & 184.00 & 181.00 & 173.00 & 151.00 & 146.00 & 143.00 & 146.00 & 163.00 \\ 
   Rook & 158.00 & 183.00 & 163.00 & 180.00 & 188.00 & 202.00 & 206.00 & 211.00 & 228.00 \\ 
   Queen & 526.00 & 480.00 & 549.00 & 606.00 & 580.00 & 558.00 & 556.00 & 540.00 & 534.00 \\ 
 \hline \\[-1.8ex] 
num observations & 92,441 & 99,319 & 94,154 & 87,957 & 82,243 & 76,453 & 68,222 & 56,997 & 44,335 \\ 
mean ply & 14.5 & 24.5 & 34.3 & 44.3 & 54.3 & 64.2 & 74.2 & 84.1 & 94 \\ 
\hline 
\hline \\[-1.8ex] 
\end{tabular} 
\end{sidewaystable} 
\begin{knitrout}\small
\definecolor{shadecolor}{rgb}{0.969, 0.969, 0.969}\color{fgcolor}\begin{figure}[h]
\includegraphics[width=0.975\textwidth,height=1.024\textwidth]{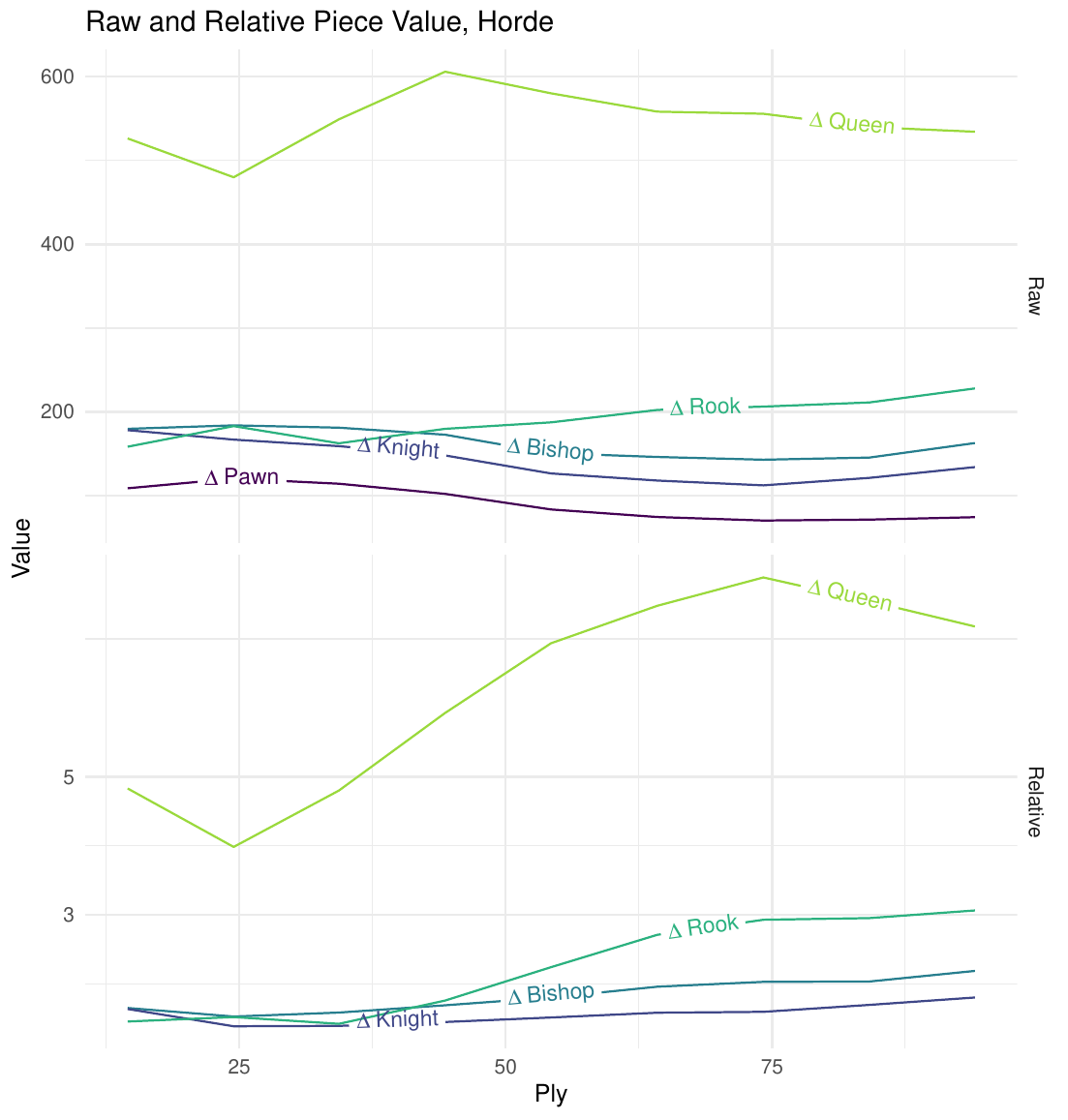} \caption[The value of piece imbalance for Horde chess is plotted versus mean ply for the ranged estimations]{The value of piece imbalance for Horde chess is plotted versus mean ply for the ranged estimations. The top facet is in raw units, while the bottom facet is relative to the absolute value of a pawn at that time. }\label{fig:horde_shorter_ranged_fit_plot}
\end{figure}

\end{knitrout}

Finally, we peel off the valuations from the earliest set of ply and
use those to quote values for equalizing games when one player is much stronger than another.
Rounding to 25 points, the
value of a pawn is approximately 100;
a knight and a bishop are both worth 175
a rook is worth 150;
and a queen is worth 525
Glicko-2 (or Elo) points.
We note that using this system to even the odds of a game is subject to any positional imbalance caused,
and of course, only major pieces can be removed from black's side.
The value of white's pawns likely depend on the file and rank, so these are only rough approximate values.


\clearpage

\section{Inferred Value from Self-Play}

Above we considered piece value by observing games between human players on Lichess.
Here we briefly examine material value as inferred from self-play of the Stockfish chess engine.  \cite{stockfish}
We set up experiments pitting two instances of Stockfish version 17.1 against each other,
starting from modified opening positions. 
We start from the standard chess opening and remove one or more pieces from one side or another,
then observe tens to thousands of matches from that starting position.
We performed a number of simulated matches from the standard opening without pieces removed, 
as well as from each of the 960 possible Chess 960 openings.
We experimented using different target Elo settings for the realizations of Stockfish, which we controlled for in our analysis.
We remove one of each of the pieces from each side, simulating hundreds of games.
We also simulated games were two major pieces are removed from one side or the other.

We nominally set the ``Elo rating'' of the stockfish engines to one of four values: 1600, 2000, 2400, 2800.
This is controlled via the \textsc{UCI\_Elo} configuration option. 
We note, however, that the ratings are calibrated to a certain time control and we did not observe the nominal spread in ratings, 
though we do see skill monotonic in the nominal ratings.
We limit play to 5 seconds per move and a total depth of 7, which is likely the limiting factor.

We perform logistic regression on the outcomes, again adjusting estimated coefficients so they are in Elo-like units.
When considering matches with different nominal strength of the two instances of Stockfish, we include terms for
engine strength in the regressions.
For these we assign a 1 to factor variables for whichever setting was used for white minus that used for black.
That is we have four variables, \textsc{Elo1600} through \textsc{Elo2800}. We initialize these four as zero, then 
increment the variable by one if that is the setting used for white, and decrement the variable corresponding to black.
Because these four variables will always sum to zero, they are not linearly independent, and we only include
\textsc{Elo2000} through \textsc{Elo2800} in the regressions, keeping \textsc{Elo1600} as the implicit ``base level''.


In \tabref{sp_mod0_standard} we report the results of three regressions:

\begin{compactenum}
\item In the first regression we consider only the standard opening, with no pieces removed.
	We include terms only for three of the four \textsc{UCI\_Elo} settings, plus a term for white's first player advantage.
	Quoted in Elo-like units, we see a spread between the settings far less than the nominal 400 point spread.
\item In the second regression we look at all standard chess matches, with and without pieces removed,
	but only consider matches where both engines used the same nominal strength rating.
	The regression only includes terms for white's advantage, and the piece imbalance.
\item In the third regression we consider all the standard chess matches we simulated.
	The regression includes terms for the engine settings, as in the first regression, white's advantage, and
	the piece imbalance.
\end{compactenum}

The engine fit values are largely consistent between the first and third regression,
and the inferred piece values are largely consistent between the second and thir regressions.
However, to our surprise, and consternation, 
the inferred piece values we fit here on self-play data are quite a bit higher than the 
``equalizer values'' we quoted above based on Lichess data.
Possible sources of this discrepancy are:
selecting non-quiescent positions in the Lichess data study;
more generally some other selection mechanism that biases our Lichess data to less decisive early material imbalances;
Stockfish's greater ability to capitalize on material differences.
We leave this question for future research.

Another surprising outcome from these regressions is the rather compressed value of the pieces:
a knight and a bishop are worth approximately 3 pawns 
(2.7 and 2.9, respectively),
while a rook is only worth around 3.6 pawns, and queen only
around 4.9 pawns. 
It is not clear why this is so compressed compared to the findings on the Lichess data or the classical piece value heuristics.

Finally we performed regressions on the full set of Standard games where we included a
one-hot encoded variable for each square holding a non-king piece in the standard opening.
This gives an inferred value for each individual piece.
As above we convert the inferred value to Elo-like units.
We plot these in \figref{standard_selfplay_plot_I}, showing the piece value at the corresponding square.
The regression is on white's probability of winning, but we flip the sign for black's pieces so they
express their value to black's chances of winning.
We observe the following:
\begin{compactenum}
\item The king and queen pawns are worth more than the knight and bishop pawns. 
	Not surprisingly the rook pawns worth less than the others, since they can only attack one other pawn.
	Black's rook pawns are worth more than white's, perhaps for their defensive value.
\item We see greater value to bishops than knights. However, there is greater value to queen side
	knights and bishops than king's side. The queen side knight is worth more than the king side bishop.
\item White's pieces have a higher fit value than black's pieces.
  Even though we control for white's first player advantage, this finding likely reflects the sensitivity
	of white's first player advantage to material imbalances.
\end{compactenum}

\begin{table}[!htbp] \centering 
  \caption{Inferred value of engine strength and material imbalance is shown via three regressions on Stockfish self-play data for standard chess. Standard errors are given in parentheses. } 
  \label{tab:sp_mod0_standard} 
\begin{tabular}{@{\extracolsep{5pt}}lccc} 
\\[-1.8ex]\hline 
\hline \\[-1.8ex] 
 & \multicolumn{3}{c}{\textit{Dependent variable:}} \\ 
\cline{2-4} 
\\[-1.8ex] & \multicolumn{3}{c}{Game Outcome} \\ 
 & All Pieces & Piece Ablations & Piece Ablations \\ 
\\[-1.8ex] & (1) & (2) & (3)\\ 
\hline \\[-1.8ex] 
 White Advantage & 16 (1) & 14 (1) & 14 (1) \\ 
  Nominal Elo 2000 & 215 (3) &  & 251 (2) \\ 
  Nominal Elo 2400 & 501 (3) &  & 520 (2) \\ 
  Nominal Elo 2800 & 657 (4) &  & 658 (3) \\ 
  $\Delta$ Pawn &  & 85 (3) & 93 (2) \\ 
  $\Delta$ Knight &  & 245 (2) & 250 (2) \\ 
  $\Delta$ Bishop &  & 265 (3) & 274 (2) \\ 
  $\Delta$ Rook &  & 321 (3) & 333 (2) \\ 
  $\Delta$ Queen &  & 455 (7) & 460 (4) \\ 
 \hline \\[-1.8ex] 
Simulated Matches & 171,000 & 180,400 & 461,000 \\ 
\hline 
\hline \\[-1.8ex] 
\end{tabular} 
\end{table} 

\begin{knitrout}\small
\definecolor{shadecolor}{rgb}{0.969, 0.969, 0.969}\color{fgcolor}\begin{figure}[h]
\includegraphics[width=0.975\textwidth,height=0.691\textwidth]{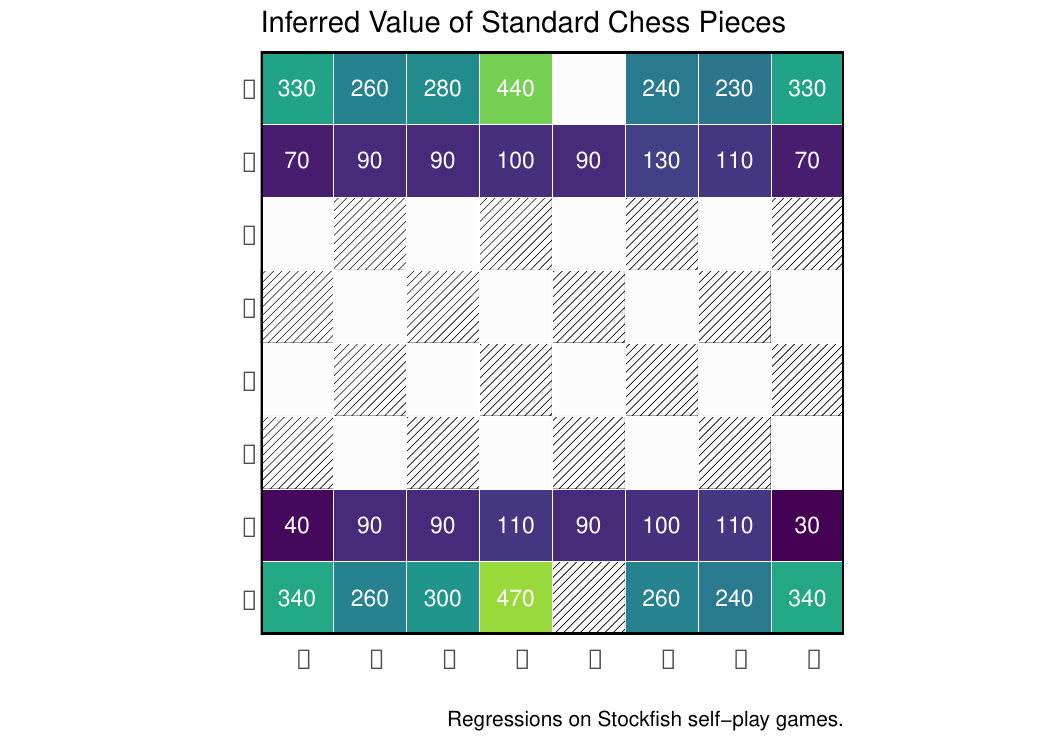} \caption[The inferred value of each piece in the standard opening are shown]{The inferred value of each piece in the standard opening are shown. Regressions on Stockfish self-play data, including all 461,000 games.}\label{fig:standard_selfplay_plot_I}
\end{figure}

\end{knitrout}

We also performed self-play experiments on each of the Chess 960 openings, with and without pieces removed.
In \tabref{sp_mod0_chess960} we report the results of three regressions, as we did for standard chess:
\begin{compactenum}
\item The first regression considers all Chess 960 openings, with no pieces removed, including
	terms only for the \textsc{UCI\_Elo} settings and white's first player advantage.
	Again the spread between settings is larger than nominal.
\item The second regression considers all Chess 960 openings, but looks only at those with equal engine settings,
	and has terms only for piece imbalance and white's first player advantage.
\item The third regression looks at all simulated matches, with terms for engine setting,
	piece imbalance and white's advantage.
\end{compactenum}

As for the simulations on standard chess, we see that 
the fit values are consistent between the first and third regression and the second and third regressions.
The fits for Chess 960 are very close to those fit from standard chess in 
\tabref{sp_mod0_standard}. Again we are surprised to find smaller value to the rook and queen, in pawn terms.
We find a knight to be worth approximately 2.6 pawns,
a bishop is worth 3.1,
a rook is worth only around 3.5 pawns, 
and a queen only around 5.6 pawns. 
Here the value of a rook is very depressed compared to classical valuation systems, worth almost as little as a bishop.
It is not clear if this is due to decreased mobility of rooks in some large proportion of Chess 960 openings.

In \figref{chess960_and_standard_ranged_fit_plot_II} we repeat the plot of 
\figref{chess960_and_standard_ranged_fit_plot}, with the estimated piece values from Lichess data for standard chess 
and Chess 960.
We add to this plot the estimated piece values for ply zero from the regressions on Stockfish self-play data.
We add a dotted line between the Lichess and Stockfish fit values.
Other than the value fit for the queen in Chess 960, the self-play fits all represent a sharp departure from the trend
seen in the Lichess fit data.
Arguably the self-play fit values are too high, or the Lichess fit values are too low for early ply ranged fits.
Further research is required to resolve these inconsistencies.

\begin{table}[!htbp] \centering 
  \caption{Inferred value of engine strength and material imbalance is shown via three regressions on Stockfish self-play data for Chess 960. Standard errors are given in parentheses. } 
  \label{tab:sp_mod0_chess960} 
\begin{tabular}{@{\extracolsep{5pt}}lccc} 
\\[-1.8ex]\hline 
\hline \\[-1.8ex] 
 & \multicolumn{3}{c}{\textit{Dependent variable:}} \\ 
\cline{2-4} 
\\[-1.8ex] & \multicolumn{3}{c}{Game Outcome} \\ 
 & All Pieces & Piece Ablations & Piece Ablations \\ 
\\[-1.8ex] & (1) & (2) & (3)\\ 
\hline \\[-1.8ex] 
 White Advantage & 15 (1) & 13 (0) & 13 (0) \\ 
  Nominal Elo 2000 & 247 (3) &  & 268 (1) \\ 
  Nominal Elo 2400 & 528 (5) &  & 530 (1) \\ 
  Nominal Elo 2800 & 688 (5) &  & 664 (1) \\ 
  $\Delta$ Pawn &  & 81 (0) & 86 (0) \\ 
  $\Delta$ Knight &  & 222 (1) & 227 (0) \\ 
  $\Delta$ Bishop &  & 262 (1) & 267 (0) \\ 
  $\Delta$ Rook &  & 290 (1) & 297 (1) \\ 
  $\Delta$ Queen &  & 479 (3) & 486 (1) \\ 
 \hline \\[-1.8ex] 
Simulated Matches & 96,000 & 1,958,400 & 4,896,000 \\ 
\hline 
\hline \\[-1.8ex] 
\end{tabular} 
\end{table} 


\begin{knitrout}\small
\definecolor{shadecolor}{rgb}{0.969, 0.969, 0.969}\color{fgcolor}\begin{figure}[h]
\includegraphics[width=0.975\textwidth,height=1.024\textwidth]{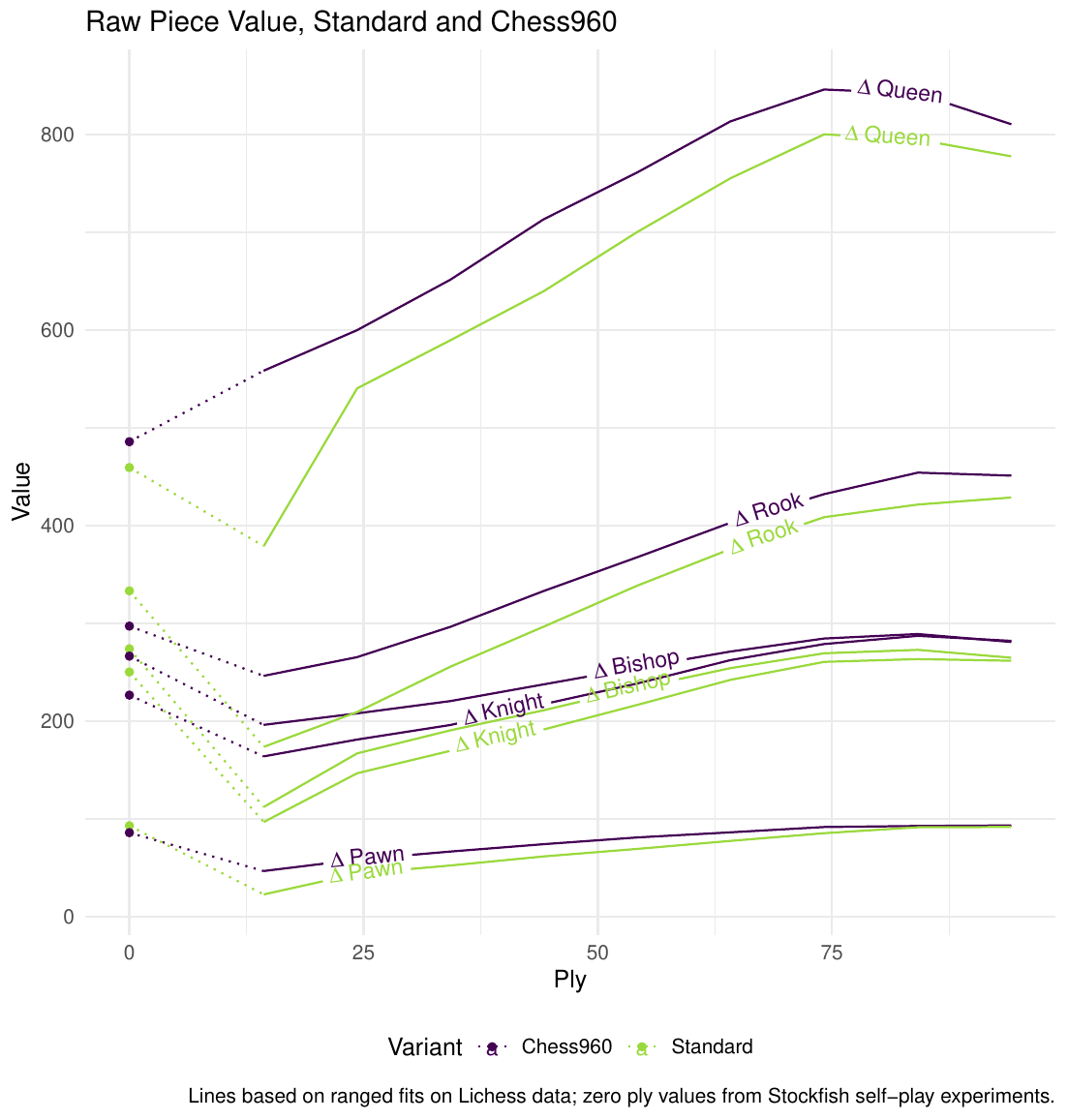} \caption[The plot of \protect\figref{chess960_and_standard_ranged_fit_plot} is augmented with values for zero ply inferred from Stockfish self play experiments]{The plot of \protect\figref{chess960_and_standard_ranged_fit_plot} is augmented with values for zero ply inferred from Stockfish self play experiments. }\label{fig:chess960_and_standard_ranged_fit_plot_II}
\end{figure}

\end{knitrout}

\section{Summary and Future Directions}

We demonstrated the use of logistic regression to estimate the value of material imbalance in Elo-like units.
Piece value as well as some positional value terms were included in various regressions on standard chess
and a few chess variants.
We found that, relative to a pawn, standard chess piece values are roughly consistent with traditional valuation systems,
but we find that bishops have slightly higher value than knights.
We found that Chess 960 exhibits valuations similar to standard chess, while Atomic chess and Antichess have very compressed valuations,
and are thus more ``positional'' than standard chess.
We find that piece values are largely negative in Antichess, with some exceptions.
We introduced a method for estimating the noise in the Glicko-2 ratings, and removing the effect of that noise from regression estimates.
We demonstrated the effect of a no-capture filter for picking out quiescent positions, though this has obvious flaws.

For future work, we foresee the following:
\begin{compactenum}
\item Obviously, a better quiescence filter is needed than the simple one chosen here. 
	That could be computationally intensive, however, and may introduce a dependence on a valuation system.
\item The differences in estimated $\sigma_0$ across the different variants is a cause for concern.
	This is a matter worth further investigation.
\item It would be interesting to see the value of more positional terms, like knights on the outer files, or having
	control of the center;
	of piece value interaction terms, like having both bishops versus a knight and a bishop, or of having both rooks;
	and of other terms, like moving a major piece twice in the opening, or having castled.
\item Our experiments with self-play raised more questions than they answered.
	Further investigation is warranted to see if piece valuations are simply different among human and engine play,
	or there is some other mechanism that explains the discrepancies we saw.
	It would be worthwhile to consider self-play experiments on positions that arise in the Lichess data,
	to compare human and machine performance.
\item We largely ignored the issue that some positions are basically winning for one side or another, while others
	are more evenly matched.
	For the lopsided positions, the difference in ratings terms should have less effect on outcomes.
	Our regressions simply compute an average value for this effect, but there should be an interaction term with the difference in rating.
	It is not clear how to construct that interaction term, moreover it would interfere with the SIMEX calibration procedure.
	Also it is likely the case that from a lopsided position, the absolute value of skill, not the difference in skill, is 
	predictive of outcomes.
	That is, a more skilled player is more likely to recognize a lopsided position in their favor and prevail from it.
	This would require additional terms in the regressions.
\item It is not clear if other adjustments should be made for snapshot ply.
	Besides the issue with lopsided positions discussed above,
	for snapshots earlier in a match there is less causal effect of skill imbalance on positional imbalance
	than one should see later in a match.
	It is not clear if any adjustments have to be made for this effect to preserve causal interpretation.
\item The SIMEX and calibration procedure is cumbersome and slow and prevents us from quoting proper confidence intervals.
	Moreover, the effect of noise in the ratings is quite a bit smaller than we found in our simulations,
	which included only a single major piece. 
	A different methodology would be welcomed.
\end{compactenum}

\nocite{savelyev:2020}
\nocite{barthelemy2024fragilitychesspositionsmeasure}
\nocite{doi:10.1177/1536867X0400300407}

\bibliographystyle{plainnat}
\bibliography{value}

@book{vanBelle2002_STRUTs,
	author = {van Belle, Gerald},
	day = {22},
	edition = {1},
	howpublished = {Paperback},
	isbn = {0471402273},
	month = "March",
	publisher = {Wiley-Interscience},
	title = {{Statistical Rules of Thumb (Wiley Series in Probability and Statistics)}},
	url = {http://www.vanbelle.org/},
	year = {2002}
}

@article{article,
author = {Bonett, Douglas},
year = {2006},
month = {02},
pages = {775-782},
title = {Approximate confidence interval for standard deviation of nonnormal distributions},
volume = {50},
journal = {Computational Statistics & Data Analysis},
doi = {10.1016/j.csda.2004.10.003}
}

@misc{silver2017masteringchessshogiselfplay,
      title={Mastering Chess and Shogi by Self-Play with a General Reinforcement Learning Algorithm}, 
      author={David Silver and Thomas Hubert and Julian Schrittwieser and Ioannis Antonoglou and Matthew Lai and Arthur Guez and Marc Lanctot and Laurent Sifre and Dharshan Kumaran and Thore Graepel and Timothy Lillicrap and Karen Simonyan and Demis Hassabis},
      year={2017},
      eprint={1712.01815},
      archivePrefix={arXiv},
      primaryClass={cs.AI},
      url={https://arxiv.org/abs/1712.01815}, 
}

@article{DBLP:journals/corr/abs-2009-04374,
  author       = {Nenad Tomasev and
                  Ulrich Paquet and
                  Demis Hassabis and
                  Vladimir Kramnik},
  title        = {Assessing Game Balance with {AlphaZero}: Exploring Alternative Rule Sets in Chess},
  journal      = {CoRR},
  volume       = {abs/2009.04374},
  year         = {2020},
  url          = {https://arxiv.org/abs/2009.04374},
  eprinttype    = {arXiv},
  eprint       = {2009.04374},
  bibsource    = {dblp computer science bibliography, https://dblp.org}
}

@Book{Levy1988,
	editor="Levy, David",
	title="Computer Chess Compendium",
	year="1988",
	publisher="Springer New York",
	address="New York, NY",
	isbn="978-1-4757-1968-0",
	doi="10.1007/978-1-4757-1968-0",
	url="https://link.springer.com/book/10.1007/978-1-4757-1968-0"
}

@InCollection{Shannon1988,
	author="Shannon, Claude E.",
	editor="Levy, David",
	title="Programming a Computer for Playing Chess",
	bookTitle="Computer Chess Compendium",
	year="1988",
	publisher="Springer New York",
	address="New York, NY",
	pages="2--13",
	doi="10.1007/978-1-4757-1968-0_2",
	url="https://doi.org/10.1007/978-1-4757-1968-0_2"
}

@InCollection{Turing1988,
	author="Turing, Alan M.",
	editor="Levy, David",
	title="Chess",
	bookTitle="Computer Chess Compendium",
	year="1988",
	publisher="Springer New York",
	address="New York, NY",
	pages="14--17",
	isbn="978-1-4757-1968-0",
	doi="10.1007/978-1-4757-1968-0_2",
	url="https://doi.org/10.1007/978-1-4757-1968-0_2"
}

@Inbook{Simon1988,
	author="Simon, Herbert and Chase, William",
	editor="Levy, David",
	title="Skill in Chess",
	bookTitle="Computer Chess Compendium",
	year="1988",
	publisher="Springer New York",
	address="New York, NY",
	pages="175--188",
	isbn="978-1-4757-1968-0",
	doi="10.1007/978-1-4757-1968-0_18",
	dumb_url="https://doi.org/10.1007/978-1-4757-1968-0_18",
	url="http://iiif.library.cmu.edu/file/Simon_box00066_fld05052_bdl0001_doc0001/Simon_box00066_fld05052_bdl0001_doc0001.pdf"
}

@incollection{birmingham1988tree,
	title={Tree-searching and tree-pruning techniques},
	author={Birmingham, John and Kent, Peter},
	bookTitle="Computer Chess Compendium",
	year="1988",
	publisher="Springer New York",
	address="New York, NY",
	pages={123--128},
	isbn="978-1-4757-1968-0",
	doi="10.1007/978-1-4757-1968-0_13",
	url="https://doi.org/10.1007/978-1-4757-1968-0_13"
}

@book{pritchard2000popular,
	title={Popular Chess Variants},
	author={Pritchard, D. B.},
	isbn={0-7134-8578-7},
	series={Batsford Chess Books},
	year={2000},
	publisher={B.T. Batsford}
}

@book{pritchard2007encyclopedia,
	title={The Classified Encyclopedia of Chess Variants},
	author={Pritchard, D. B.},
	note={The second edition of The Encyclopedia of Chess Variants},
	isbn={978-0-9555168-0-1},
	publisher={John Beasley},
	url={https://www.jsbeasley.co.uk/encyc/encyc.pdf},
	year={2007}
}

@book{andrejic2018ultimate,
	title={The Ultimate Guide to Antichess},
	author={Andreji\'{c}, Vladica},
	publisher={\v{S}ahovski informator},
	url={https://www.sahovski.com},
	isbn={978-8-7297-09-8},
	year={2018}
}

@misc{enwiki:1265673175,
	author = "{Wikipedia contributors}",
	title = "Chess piece relative value --- {Wikipedia}{,} The Free Encyclopedia",
	year = "2024",
	url = "https://en.wikipedia.org/w/index.php?title=Chess_piece_relative_value&oldid=1265673175",
	note = "[Online; accessed 28-December-2024]"
}

@misc{betza1996,
	author={Betza, Ralph},
	title={About the Values of Chess Pieces},
	url={https://www.chessvariants.com/d.betza/pieceval/index.html},
	year=1996
}

@article{kaufman1999evaluation,
	title={The evaluation of material imbalances},
	author={Kaufman, Larry},
	journal={Chess Life},
	url={https://archive.ph/20120714204040/http://mywebpages.comcast.net/danheisman/Articles/evaluation_of_material_imbalance.htm},
	year={1999}
}

@article{glickman1995glicko,
	title={The Glicko system},
	author={Glickman, Mark E},
	journal={Boston University},
	volume={16},
	number={8},
	pages={9},
	url={http://www.fngo.it/public/pdf/glicko.pdf},
	year={1995}
}

@book{elo1978rating,
	title={The Rating of Chessplayers: Past and Present},
	author={Elo, Arpad E.},
	isbn={9780923891275},
	lccn={2010549499},
	year={1978},
	publisher={Ishi Press International}
}

@article{elo1967proposal,
	author={Elo, Arpad E.},
	title={The Proposed {USCF} Rating System, Its Development, Theory, and Applications},
	journal={Chess Life},
	volume={XXII},
	number=8,
	month={August},
	page={242-247},
	year=1967
}

@article{gupta2023,
	author={Gupta, Aditya and Gupta, Arnav and Grattoni, Christopher},
	title={Determining Chess Piece Values Using Machine Learning},
	journal={Journal of Student Research},
	doi={10.47611/jsrhs.v12i1.4356},
	url={https://www.jsr.org/hs/index.php/path/article/view/4356/1910},
	volume=12,
	number=1,
	year=2023
}

@misc{winter2012,
	author={Winter, Edward},
	title={The Value of the Chess Pieces},
	url={https://www.chesshistory.com/winter/extra/value.html},
	year={2012}
}

@book{maizelis2014soviet,
	title={The {S}oviet Chess Primer},
	author={Maizelis, Ilya Lvovich},
	isbn={978-1-907982-99-6},
	series={Chess classics},
	year={2014},
	publisher={Quality Chess}
}

@article{watkins2017losing,
	title={Losing Chess: 1. {e3} Wins for White},
	author={Watkins, Mark},
	journal={J. Int. Comput. Games Assoc.},
	volume={39},
	number={2},
	pages={123--125},
	url={https://magma.maths.usyd.edu.au/~watkins/LOSING_CHESS/LCsolved.pdf},
	year={2017}
}

@mastersthesis{savelyev:2020,
	author={Savelyev, Sergey},
	title={Mastering Reconnaissance Blind Chess with Reinforcement Learning},
	type={Bachelor's Thesis},
	url={http://hdl.handle.net/1853/63890},
	school={Georgia Institute of Technology},
	month={May},
	year=2020
}

@misc{barthelemy2024fragilitychesspositionsmeasure,
	title={Fragility of Chess positions: measure, universality and tipping points},
	author={Marc Barthelemy},
	year={2024},
	eprint={2410.02333},
	archivePrefix={arXiv},
	primaryClass={physics.soc-ph},
	url={https://arxiv.org/abs/2410.02333}
}

@book{standage2002turk,
	title={The {T}urk: The life and times of the famous eighteenth-century chess-playing machine},
	author={Standage, Tom},
	year={2002},
	isbn={978-0-8027-1391-9},
	publisher={Walker \& Company}
}

@misc{enwiki:1270044935,
	author = "{Wikipedia contributors}",
	title = "Mechanical Turk --- {Wikipedia}{,} The Free Encyclopedia",
	year = "2025",
	url = "https://en.wikipedia.org/w/index.php?title=Mechanical_Turk&oldid=1270044935",
	note = "[Online; accessed 23-January-2025]"
}

@misc{enwiki:1261068408,
	author = "{Wikipedia contributors}",
	title = "{D}eep {B}lue versus {G}arry {K}asparov --- {Wikipedia}{,} The Free Encyclopedia",
	year = "2024",
	url = "https://en.wikipedia.org/w/index.php?title=Deep_Blue_versus_Garry_Kasparov&oldid=1261068408",
	note = "[Online; accessed 23-January-2025]"
}

@book{pandolfini1997kasparov,
	title={Kasparov and Deep Blue: The historic chess match between man and machine},
	author={Pandolfini, Bruce},
	year={1997},
	publisher={Simon and Schuster}
}

@inproceedings{chitale2024implementing,
  author={Chitale, Angad Makarand and Cherian, Aaron Mano and Singh, Amitoz and Prakasam, P},
  booktitle={2024 IEEE International Conference on Electronics, Computing and Communication Technologies (CONECCT)},
	title={Implementing the Chess Engine using {NNUE} with Nega-Max Algorithm},
  year={2024},
  volume={},
  number={},
  pages={1-6},
	organization={IEEE},
	url={https://ieeexplore.ieee.org/abstract/document/10677087},
  doi={10.1109/CONECCT62155.2024.10677087}
}

@misc{klein2022neuralnetworkschess,
	title={Neural Networks for Chess}, 
	author={Dominik Klein},
	year={2022},
	eprint={2209.01506},
	archivePrefix={arXiv},
	primaryClass={cs.LG},
	url={https://arxiv.org/abs/2209.01506}, 
}

@book{lasker1926,
  title={Lasker's manual of chess},
  author={Lasker, Emanuel},
  year={1926},
	url={http://simardartizanfarm.ca/pdf/-_Lasker_s_Manual_of_Chess.pdf}
}

@book{fine1942chess,
  title={Chess the Easy Way},
  author={Fine, Reuben},
  lccn={42009908},
  series={Cornerstone Library books},
  year={1942},
  publisher={David McKay Company}
}

@misc{lichess,
  title = {Lichess},
	year="2010",
  url = {https://lichess.org}
}

@misc{stockfish,
  title = {Stockfish},
	year="2025",
  url = {https://stockfishchess.org}
}

@misc{gilgamath-atomic-two,
	author={Pav, Steven},
	title={Atomic Piece Values},
	url={https://www.gilgamath.com/atomic-two},
	year={2021}
}

@misc{gilgamath-atomic-three,
	author={Pav, Steven},
	title={Atomic Piece Values, Again},
	url={https://www.gilgamath.com/atomic-three},
	year={2021}
}

@book{mccullagh2019generalized,
  title={Generalized linear models},
  author={McCullagh, Peter and Nelder, J. A.},
  publisher={Routledge},
	doi={10.1201/9780203753736},
	url={https://doi.org/10.1201/9780203753736},
  year={1989}
}

@book{pearl2018book,
  title={The book of why: the new science of cause and effect},
  author={Pearl, Judea and Mackenzie, Dana},
  year={2018},
  publisher={Basic books}
}

@book{angrist2009mostly,
  title={Mostly harmless econometrics: An empiricist's companion},
  author={Angrist, Joshua D and Pischke, J{\"o}rn-Steffen},
  year={2009},
  publisher={Princeton university press}
}

@misc{hernan2020causal,
  title={Causal Inference: What if},
	publisher={Chapman Hill/CRC},
  author={Hern{\'a}n, MA and Robins, JM},
	url={https://miguelhernan.org/whatifbook},
  year={2020}
}

@book{scharnagl2004fischer,
  title={Fischer-Random-Schach: (FRC/Chess 960) ; die revolution{\"a}re Zukunft des Schachspiels (incl. Computerschach)},
  author={Scharnagl, Reinhard},
  isbn={978-3-8334-1322-3},
  year={2004},
  publisher={Books on Demand}
}

@article{doi:10.1177/1536867X0400300407,
author = {James W. Hardin and Henrik Schmiediche and Raymond J. Carroll},
title ={The Simulation Extrapolation Method for Fitting Generalized Linear Models with Additive Measurement Error},
journal = {The Stata Journal},
volume = {3},
number = {4},
pages = {373-385},
year = {2003},
doi = {10.1177/1536867X0400300407},
url = {https://doi.org/10.1177/1536867X0400300407},
eprint = {https://doi.org/10.1177/1536867X0400300407}
}

@article{fujii2013extended,
  title={Extended {K}alman filter},
  author={Fujii, Keisuke},
  journal={Refernce Manual},
  volume={14},
  pages={41},
	url={https://www-jlc.kek.jp/2004sep/subg/offl/kaltest/doc/ReferenceManual.pdf},
  year={2013}
}

@book{kennedy2003guide,
  title={A Guide to Econometrics},
  author={Kennedy, Peter},
  isbn={9780262611831},
  lccn={20045179},
  series={Mit Press},
  year={2003},
  publisher={MIT Press}
}

@book{enigma2021definitive,
  title={The Definitive Guide to Horde Chess: Openings, Strategies and Tactics for White and Black},
  author={Enigma Chess},
  isbn={9798737259457},
  series={Chess Variants Theory Books},
  url={https://books.google.com/books?id=9npyzgEACAAJ},
  year={2021},
  publisher={Amazon Digital Services LLC - KDP Print US}
}

@misc{szczecinski2021simplifiedkalmanfilteronline,
      title={Simplified {K}alman filter for online rating: one-fits-all approach}, 
      author={Leszek Szczecinski and Rapha{\"{e}}lle Tihon},
      year={2021},
      eprint={2104.14012},
      archivePrefix={arXiv},
      primaryClass={stat.ML},
      url={https://arxiv.org/abs/2104.14012}, 
}

@book{carroll1995measurement,
  title={Measurement error in nonlinear models},
  author={Carroll, Raymond J. and Ruppert, David and Stefanski, Leonard A.},
  volume={105},
  year={1995},
  dumburl={http://ndl.ethernet.edu.et/bitstream/123456789/88802/1/CarrollRupertStefanskiCrainiceanu2006%20measurement%20error%20in%20nonlinear%20models.pdf},
  publisher={CRC press}
}

@article{hardin2003regression,
  title={The regression-calibration method for fitting generalized linear models with additive measurement error},
  author={Hardin, James W and Schmiediche, Henrik and Carroll, Raymond J.},
  journal={The Stata Journal},
  volume={3},
  number={4},
  pages={361--372},
  year={2003},
  url={https://www.stata.com/merror/rcal.pdf},
  publisher={SAGE Publications Sage CA: Los Angeles, CA}
}

@book{efron1994,
	title={An Introduction to the Bootstrap},
	author={Efron, Bradley and Tibshirani, R. J.},
	publisher={Chapman and Hall/CRC},
	doi={10.1201/9780429246593},
	url={https://doi.org/10.1201/9780429246593},
	year={1994}
}

@book{dobson_introduction_2008,
	title = {An Introduction to Generalized Linear Models},
	isbn = {978-1-58488-951-9},
	language = {en},
	publisher = {CRC Press},
	author = {Dobson, Annette J. and Barnett, Adrian},
	month = may,
	year = {2008},
	note = {Google-Books-ID: ZzncBQAAQBAJ},
}

\appendix

\section{Standard Error of Rating Scores}
\label{sec:elo_se}

Suppose two players repeatedly play a chess-like game, and we observe the outcome, as a win, loss or draw,
and update a skill rating for each player.
Let $\Delta a$ be the difference in true underlying ability of the players, and
let $\Delta r$ be the difference in their observed rating at some point in time.
Both these differences are with respect to the frist player.
Let $y$ be the outcome of one game, a $1$ if first player wins, $1/2$ in case of a draw, and $0$ if first player loses.
After observing the outcome $y$ we increase the first player's rating by some amount (positive or negative), 
and decrease the second player's rating by the same amount.
We note this is not exactly how the Glicko-2 score is updated, as players can have asymmetric uncertainties, 
but is close to accurate if the players have similar histories.
The difference in ratings thus is changed by twice that amount. We write this as
\begin{equation}
	\Delta r \leftarrow \Delta r + 2 k \wrapParens{y - \funcit{g}{\Delta r}},
\end{equation}
where $k$ is the $k$-factor (typically in the range of 10 to 40), and
the function $\funcit{g}{\cdot}$ is the expected value of $y$ based on the current estimate of the difference in abilities:
\begin{equation}
	\funcit{g}{\Delta r} 
	= \frac{10^{\Delta r / 400}}{1 + 10^{\Delta r / 400}} 
	= \frac{\exp{\cfac \Delta r}}{1 + \exp{\cfac \Delta r}}.
\end{equation}

Let $\epsilon$ be the error between the ability and rating values: $\Delta r = \Delta a + \epsilon$.
Then the error is updated as
\begin{equation}
	\epsilon \leftarrow \epsilon + 2 k \wrapParens{y - \funcit{g}{\Delta a + \epsilon}}.
\end{equation}
Use Taylor's theorem to linearize the function $\funcit{g}{\cdot}$ around $\Delta a$ to get approximately:
\begin{align}
\epsilon 
	\nonumber
	&\overset{\approx}{\leftarrow} \epsilon + 2 k \wrapParens{y - \funcit{g}{\Delta a} - \epsilon \funcit{g'}{\Delta a}},\\
	&\, = 
	2 k \wrapParens{y - \funcit{g}{\Delta a}} + \wrapParens{1  - 2k \funcit{g'}{\Delta a}}\epsilon.
	\label{eqn:epsi_update}
\end{align}

We note that 
\begin{equation}
	\funcit{g'}{\Delta a} = \cfac \funcit{g}{\Delta a}\wrapParens{1 - \funcit{g}{\Delta a}}.
\end{equation}
Now we note that $\abs{\funcit{g'}{\Delta a}} \le \cfac$.
Thus if $k \cfac < 1$, or equivalently, $k < 173.72$,
then $\abs{1  - 2k \funcit{g'}{\Delta a}} < 1$ 
and \eqnref{epsi_update} defines an AR(1) process on $\epsilon$ that is stationary in the weak sense.
This is due to the fact that the term $\wrapParens{y - \funcit{g}{\Delta a}}$ is zero mean.

The asymptotic variance of $\epsilon$ will then be
\begin{align}
	\VAR{\epsilon}
	\nonumber
	&= \frac{4k^2 \VAR{y - \funcit{g}{\Delta a}}}{1 - {\wrapParens{1  - 2k \funcit{g'}{\Delta a}}}^2},\\
	\nonumber
	&= \frac{4k^2 \VAR{y - \funcit{g}{\Delta a}}}{4k \funcit{g'}{\Delta a} - 4k^2 \wrapParens{\funcit{g'}{\Delta a}}^2},\\
	\nonumber
	&= \frac{k \VAR{y - \funcit{g}{\Delta a}}}{\funcit{g'}{\Delta a} - k \wrapParens{\funcit{g'}{\Delta a}}^2},\\
	\label{eqn:var_epsilon_I}
	&= \frac{k \sigma^2}{\funcit{g'}{\Delta a} \wrapParens{1 - k \funcit{g'}{\Delta a}}},
\end{align}
where $\sigma^2$ is the variance of $y$.

We wish to use \eqnref{var_epsilon_I} to compute the standard error of $\Delta r$.
First we compute $\sigma^2$ by parametrizing the probability of a draw, which we denote as $p$.
As an aside, $p$ is generally observed to be larger when absolute skill is higher,
with more than half of high-level chess tournament games ending in a draw.
To get the expected values to work out, we will observe $y=1$ with probability
$\funcit{g}{\Delta a} - p$. 
The variance of $y$ is then
\begin{equation}
	\sigma^2 = \funcit{g}{\Delta a}\wrapParens{1 - \funcit{g}{\Delta a}} - \frac{p}{4}.
\end{equation}

\begin{knitrout}\small
\definecolor{shadecolor}{rgb}{0.969, 0.969, 0.969}\color{fgcolor}\begin{figure}[h]
\includegraphics[width=0.975\textwidth,height=0.691\textwidth]{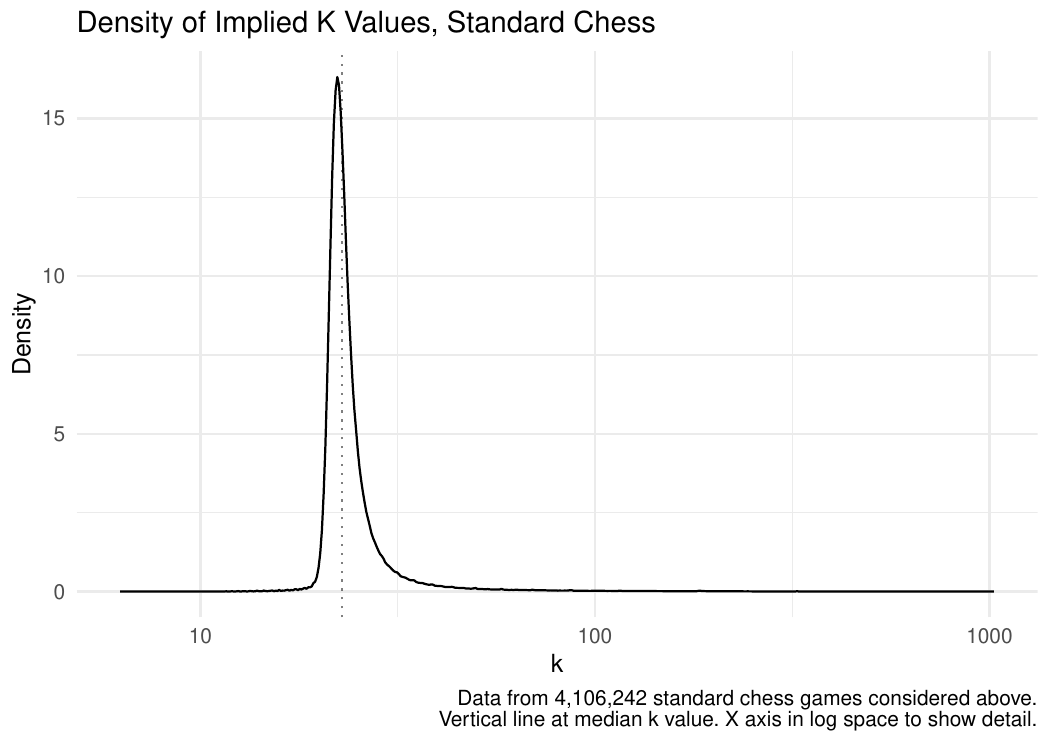} \caption[The density of implied $k$ values from the 4,106,242 standard chess games considered in \secref{regression_analysis} is plotted]{The density of implied $k$ values from the 4,106,242 standard chess games considered in \secref{regression_analysis} is plotted. A horizontal line is plotted at the median value, $22.83$.}\label{fig:elo_kvalue_plot}
\end{figure}

\end{knitrout}

To get some idea of the value of $k$, we consider the 4,106,242 standard chess games
considered in \secref{regression_analysis}, with all the filters applied there.
We observed the difference in Glicko-2 scores before and after each game, as well as the outcome
of the game. 
From these we can back out a value of $k$ for each game.
We plot all these $k$ values in \figref{elo_kvalue_plot}, and compute the median value as $22.83$.

We now compute the standard error of $\Delta r$ from \eqnref{var_epsilon_I}
for various values of $\Delta a$, $k$, and $p$.
We plot these in \figref{elo_se_lines}.
We plot a vertical line at the median value of implied $k$,
and a horizontal line at the standard error in ratings that we estimated from the SIMEX procedure in 
\secref{regression_analysis}, $58$.
We find that value of standard error is entirely consistent with our analysis,
though it would imply a somewhat low probability of a tie.
We note, however, that our analysis is based on a linear approximation to the update formula,
that a whole distribution of $k$ values is used, including values larger than our median,
and that our analysis assumes that $\Delta a$ is constant over time.

\begin{knitrout}\small
\definecolor{shadecolor}{rgb}{0.969, 0.969, 0.969}\color{fgcolor}\begin{figure}[h]
\includegraphics[width=0.975\textwidth,height=0.691\textwidth]{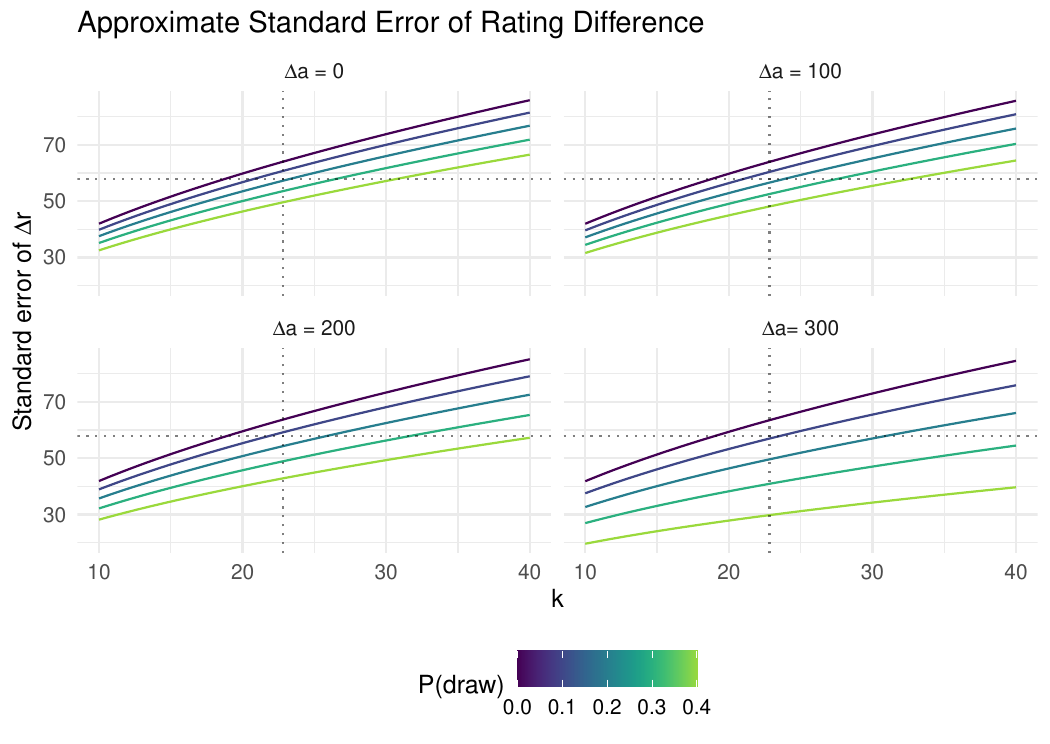} \caption[The standard error of $\Delta r$ implied by \eqnref{var_epsilon_I} is plotted versus $k$ for various values of $p$ and $\Delta a$]{The standard error of $\Delta r$ implied by \eqnref{var_epsilon_I} is plotted versus $k$ for various values of $p$ and $\Delta a$. We plot a vertical line at the median value of $k$ from the sample of standard games, $22.83$ and a horizontal line at the implied standard error based on the SIMEX procedure, $58$. }\label{fig:elo_se_lines}
\end{figure}

\end{knitrout}

\clearpage

\section{Regression Calibration with a Single Noisy Covariate}
\label{sec:reg_cal}

Here we consider the regression calibration method when a single covariate is observed with noise.
Suppose that vector $\vect{x}$ represents a true covariate which we wish to use in a regression.
However we instead observe $\hatvect{x}$, where $x_1$ has some added zero-mean noise with standard deviation $\sigma$.
One way to estimate the true regression effects against $\vect{x}$ is to 
translate the results from a regression against $\hatvect{x}$.

So suppose we observe some outcome $y$ with
$$
\funcit{f}{y} \sim \trvect{\beta}\vect{x},
$$
but we can only estimate $\vect{\beta}$ by regression against the observable $\hatvect{x}$, yielding estimate
$\hatvect{\beta}$.
We can estimate the relationship between $\vect{\beta}$ and $\hatvect{\beta}$ by imagining a linear relationship
$$
\vect{x} \approx \Mtx{A} \hatvect{x},
$$
whereupon we then write
$$
\funcit{f}{y} \sim \trvect{\beta}\vect{x} \approx \trvect{\beta}\Mtx{A}\hatvect{x} = \tr{\hatvect{\beta}}\hatvect{x}.
$$

We estimate the matrix \Mtx{A} by a linear regression of \hatvect{x} on \vect{x}.
This is the method of \emph{regression calibration}.  \cite{carroll1995measurement,hardin2003regression}
Suppose one observes $\hatMtx{X}$ whose rows are independent observations of the vectors $\hatvect{x}$.
Similarly let $\Mtx{X}$ be the matrix whose rows are the corresponding \vect{x}.
The regression calibration is
\begin{equation}
	{\hatvect{\beta}} \approx \minvParens{\tr{\hatMtx{X}}\hatMtx{X}} \wrapParens{\tr{\hatMtx{X}}\Mtx{X}} {\vect{\beta}}.
\end{equation}
While this makes the most sense when the response $y$ is linear in $\trvect{\beta}\vect{x}$, it
is used in the GLM context as well.

We now write
\begin{equation}
\hatMtx{X} = \Mtx{X} + \sqrt{n}\sigma\vect{z}\trbasev[1],
\end{equation}
where $\vect{z}$ is a vector of sample disturbances.
Let us also assume that
\begin{align}
	\trvect{z}\Mtx{X} &\approx \tr{\vzero},\\
	\trvect{z}\vect{z} &= 1.
\end{align}
These are respectively that $\vect{z}$ is orthogonal, or independent, from the columns of \Mtx{X} and calibrated to have unit standard deviation.

Then
$$
\wrapParens{\tr{\hatMtx{X}}\hatMtx{X}} \approx \tr{\Mtx{X}}\Mtx{X} + n\sigma^2\basev[1]\trbasev[1].
$$
Via the Sherman-Morrison-Woodbury formula, then
$$
\minvParens{\tr{\hatMtx{X}}\hatMtx{X}} \approx \minvParens{\tr{\Mtx{X}}\Mtx{X}} - \frac{n\sigma^2 \minvParens{\tr{\hatMtx{X}}\hatMtx{X}}\basev[1]\trbasev[1]\minvParens{\tr{\hatMtx{X}}\hatMtx{X}}}{1 + n\sigma^2\trbasev[1] \minvParens{\tr{\hatMtx{X}}\hatMtx{X}} \basev[1]}.
$$
By assumption we have
$$
\wrapParens{\tr{\hatMtx{X}}\Mtx{X}} \approx \tr{\Mtx{X}}\Mtx{X},
$$
and so
\begin{align*}
	\minvParens{\tr{\hatMtx{X}}\hatMtx{X}} \wrapParens{\tr{\hatMtx{X}}\Mtx{X}} \vect{\beta} 
	&\approx 
	\vect{\beta} - \frac{n\sigma^2 \minvParens{\tr{\hatMtx{X}}\hatMtx{X}}\basev[1]\trbasev[1]\vect{\beta}}{1 + n\sigma^2\trbasev[1] \minvParens{\tr{\hatMtx{X}}\hatMtx{X}} \basev[1]}.
\end{align*}
The first element of this, the coefficient associated with the noisily observed variable, takes value
\begin{align*}
	\hat{\beta}_1 &\approx \trbasev[1]\minvParens{\tr{\hatMtx{X}}\hatMtx{X}} \wrapParens{\tr{\hatMtx{X}}\Mtx{X}} \vect{\beta}\\
	&\approx 
	\trbasev[1]\vect{\beta} - \frac{n\sigma^2\trbasev[1]\minvParens{\tr{\hatMtx{X}}\hatMtx{X}}\basev[1]\trbasev[1]\vect{\beta}}{1 + n\sigma^2\trbasev[1] \minvParens{\tr{\hatMtx{X}}\hatMtx{X}} \basev[1]},\\
	&=
	\frac{1}{1 + n\sigma^2\trbasev[1] \minvParens{\tr{\hatMtx{X}}\hatMtx{X}} \basev[1]}\trbasev[1]\vect{\beta},\\
	&=
	\frac{\beta_1}{1 + c\sigma^2}.
\end{align*}

The other elements have values like:

\begin{align*}
	\hat{\beta}_2 &\approx \trbasev[2]\minvParens{\tr{\hatMtx{X}}\hatMtx{X}} \wrapParens{\tr{\hatMtx{X}}\Mtx{X}} \vect{\beta}\\
	&\approx 
	\trbasev[2]\vect{\beta} - \frac{n\sigma^2\trbasev[2]\minvParens{\tr{\hatMtx{X}}\hatMtx{X}}\basev[1]\trbasev[1]\vect{\beta}}{1 + n\sigma^2\trbasev[1] \minvParens{\tr{\hatMtx{X}}\hatMtx{X}} \basev[1]},\\
	&=\beta_2 - \frac{k\sigma^2 \beta_1}{1 + c\sigma^2},\\
	&=\beta_2 - k\sigma^2 \hat{\beta}_1.
\end{align*}

Together these suggest that one can estimate $\beta_1$ by observing $\hat{\beta}_1$ for multiple values of $\sigma$ via SIMEX,
then regressing $\hat{\beta}_1^{-1}$ against $\sigma^2$, then inverting the intercept term.
To estimate \emph{e.g.} $\beta_2$, compute $\hat{\beta}_1, \hat{\beta}_2$ for multiple values of $\sigma$ via SIMEX,
then regress $\hat{\beta}_2$ against $\sigma^2 \hat{\beta}_1$.


\section{Alternative Valuations}

In \tabref{value_systems}, we present some historical valuations systems,
as collected by Wikipedia editors and \citeauthor{winter2012}. \cite{enwiki:1265673175,winter2012}
The ``Mobility'' entry is based on the average mobility of pieces and assuming a pawn value of 1.75.

\begin{table}[ht]
\centering
\begin{tabular}{llrrrr}
  \hline
source & year & knight & bishop & rook & queen \\ 
  \hline
Mobility &  & 3.00 & 5.00 & 8.00 & 13.00 \\ 
  Modenese & 1750 & 3.00 & 3.00 & 5.00 & 9.00 \\ 
  Sarratt & 1813 & 3.10 & 3.30 & 5.00 & 7.90 \\ 
  Philidor & 1817 & 3.05 & 3.50 & 5.48 & 9.94 \\ 
  Peter Pratt & 1833 & 3.00 & 3.00 & 5.00 & 10.00 \\ 
  Bilguer & 1843 & 3.50 & 3.50 & 5.70 & 10.30 \\ 
  Tomlinson & 1845 & 3.05 & 3.50 & 5.48 & 9.94 \\ 
  Lasker & 1934 & 3.00 & 3.00 & 5.00 & 9.50 \\ 
  Maizelis & 1936 & 3.50 & 3.50 & 5.00 & 9.75 \\ 
  Fine & 1942 & 3.00 & 3.00 & 5.00 & 9.00 \\ 
  Euwe & 1944 & 3.50 & 3.50 & 5.50 & 10.00 \\ 
  Lasker & 1947 & 3.50 & 3.50 & 5.00 & 8.50 \\ 
  Horowitz & 1951 & 3.00 & 3.10 & 5.00 & 9.00 \\ 
  Turing & 1953 & 3.00 & 3.50 & 5.00 & 10.00 \\ 
  Evans & 1958 & 3.50 & 3.62 & 5.00 & 10.00 \\ 
  Styeklov & 1961 & 3.50 & 3.50 & 5.00 & 9.50 \\ 
  Fischer & 1972 & 3.00 & 3.25 & 5.00 & 9.00 \\ 
  Euwe & 1974 & 3.00 & 3.00 & 4.25 & 8.50 \\ 
  Kasparov & 1986 & 3.00 & 3.15 & 4.50 & 9.00 \\ 
  Soviet chess encyclopedia & 1990 & 3.00 & 3.00 & 5.00 & 9.50 \\ 
  Hooper and Whyld & 1992 & 4.00 & 3.50 & 7.00 & 13.50 \\ 
  Berliner & 1999 & 3.20 & 3.33 & 5.10 & 8.80 \\ 
  Kaufman & 1999 & 3.25 & 3.25 & 5.00 & 9.75 \\ 
  Kaufman & 2011 & 3.50 & 3.50 & 5.25 & 10.00 \\ 
  Kurzdorfer & 2003 & 3.50 & 3.50 & 5.00 & 9.00 \\ 
  Soltis & 2004 & 3.00 & 3.00 & 4.50 & 9.00 \\ 
  Yevgeny Gik & 2004 & 2.40 & 4.00 & 6.40 & 10.40 \\ 
  AlphaZero & 2020 & 3.05 & 3.33 & 5.63 & 9.50 \\ 
   \hline
\end{tabular}
\caption{Various value systems are shown, implicitly relative to the value of a pawn. Data from Wikipedia and \protect\citeauthor{winter2012}. \protect\cite{enwiki:1265673175,winter2012}} 
\label{tab:value_systems}
\end{table}

\end{document}